\documentclass[aps,preprint,amsmath,amssymb,t1,11pt]{revtex4}
\usepackage{graphicx}
\usepackage[T1]{fontenc}
\usepackage[a4paper,margin=2.5cm]{geometry}
\usepackage{physics}
\usepackage{bm}
\usepackage{booktabs}
\usepackage{setspace}
\usepackage{titlesec}
\usepackage{caption}
\usepackage{fancyhdr}
\usepackage{mathtools}
\usepackage{enumitem}
\usepackage{hyperref}
\hypersetup{
    colorlinks=true,
    linkcolor=blue,
    filecolor=magenta,      
    urlcolor=cyan,
}

\begin{document}

\title{Ground-State Energy Solutions of the Lithium Atom: Zeroth-, First-, and Second-Order Perturbation Theory and the Variational Method}

\author{Afraa Mahboubi}\email[]{m.afra11211@gmail.com}
\author{Büşra Gökçe Zolmaz}\email[]{busra200202024@gmail.com}
\author{Devrim Karayer}\email[]{devrimkarayer@gmail.com}
\author{Handenur Şay}\email[]{hndnrsy555@gmail.com}
\author{Oğuzhan Kaya}\email[]{oguzhan.ky98@gmail.com}
\author{Abdulkadir Senol}\email[]{senol\_a@ibu.edu.tr}
\affiliation{Department of Physics, Bolu Abant Izzet Baysal University, 14280, Bolu, T\"{u}rkiye}
\date{\today}

\begin{abstract}
In this work, the ground-state energy of the lithium atom is systematically investigated using both time-independent perturbation theory and the variational method to provide a comprehensive pedagogical analysis of many-body atomic systems. The unperturbed Hamiltonian is initially constructed by neglecting electron-electron interactions, treating the system as three independent hydrogen-like electrons to yield a zeroth-order energy baseline of -275.51 eV. The antisymmetric fermionic nature of the exact wave function is rigorously enforced through the Slater determinant formalism. First-order perturbation theory is applied to evaluate static inter-electronic repulsion using exact Coulomb and exchange integrals, refining the energy state to -192.01 eV. To account for dynamical electronic correlation, second-order perturbation theory is computed numerically for virtual single-electron s-orbital transitions, leading to a total perturbative energy of -196.36 eV. A brief discussion of two-electron excitations is also included to encapsulate further physical realism within the framework. Furthermore, a non-orthogonal two-parameter variational approach is employed to model the shell-specific shielding effect. By optimizing the effective nuclear charges, the variational method establishes a superior upper bound energy of -201.187 eV. The results of both methods are comprehensively contrasted against each other and the reference baseline to provide critical insights into the nature of electron correlation and screening in multi-electron atoms.
\end{abstract}
\maketitle
\section{Introduction}
The lithium atom represents the simplest multi-electron atomic system beyond helium that eludes an exact analytical solution. Unlike the hydrogen atom, for which the Schrödinger equation yields exact closed-form solutions, the presence of electron-electron repulsion in lithium introduces dynamic correlation effects that necessitate approximation methods. Consequently, techniques such as perturbation theory and the variational method serve as indispensable tools in atomic quantum mechanics.

In this study, both approaches are systematically employed and contrasted with each other, as well as with reference benchmarks. This comparative analysis serves a pedagogical purpose by illustrating the limitations of the independent-particle approximation and highlighting the quantitative and conceptual significance of electron correlations in multi-electron systems. Throughout this work, time-independent perturbation theory and the variational method are applied within the framework of the Born-Oppenheimer approximation, where the nucleus is treated as stationary. The zeroth-order perturbative baseline treats the electrons as independent particles moving within the nuclear Coulomb potential, while inter-electronic repulsion terms are subsequently introduced as perturbative corrections. The variational method on the other hand utilizes a trial function with two varied  parameters, each representing the shielding effect on the potential energy in each orbital, and than minimizing the expectation value of the Hamiltonian in order to estimate the ground state energy.

The ground-state electronic configuration of the lithium atom is given by
\[
1s^2 2s^1
\]
which consists of two electrons occupying the inner $1s$ shell and a single valence electron in the $2s$ orbital.

\section{Non-Relativistic Hamiltonian of the Lithium Atom}
The time-independent Schrödinger equation is expressed as \cite[Sec. 5.2]{griffiths2018}:
\begin{equation}
\hat{H}\Psi = E\Psi
\end{equation}

For a lithium atom with three electrons, the exact non-relativistic Hamiltonian ($\hat{H}$), explicitly expanding all individual kinetic and potential energy components, is given by \cite[Sec. 5.2]{griffiths2018}:
\begin{equation}
\hat{H} = \sum_{i=1}^{3} \left[ -\frac{\hbar^2}{2m}\nabla_i^2 - \frac{Ze^2}{4\pi\varepsilon_0 r_i} \right] + \sum_{i<j}^{3} \frac{e^2}{4\pi\varepsilon_0 r_{ij}}
\end{equation}
where $Z = 3$ represents the atomic number of lithium, $r_i$ denotes the distance between the nucleus and the $i$-th electron, and $r_{ij} = |\mathbf{r}_i - \mathbf{r}_j|$ represents the relative inter-electronic distance between electrons $i$ and $j$.
Expanding $\hat{H}$ term-by-term into individual electronic coordinates yields:
\begin{equation}
\begin{aligned}
\hat{H} &= \left[ -\frac{\hbar^2}{2m}\nabla_1^2 - \frac{3e^2}{4\pi\varepsilon_0 r_1} \right] + \left[ -\frac{\hbar^2}{2m}\nabla_2^2 - \frac{3e^2}{4\pi\varepsilon_0 r_2} \right] \\
&\quad + \left[ -\frac{\hbar^2}{2m}\nabla_3^2 - \frac{3e^2}{4\pi\varepsilon_0 r_3} \right] + \frac{e^2}{4\pi\varepsilon_0 r_{12}} + \frac{e^2}{4\pi\varepsilon_0 r_{13}} + \frac{e^2}{4\pi\varepsilon_0 r_{23}}
\end{aligned}
\end{equation}

The coupled nature of the inter-electronic repulsion terms ($1/r_{ij}$) mathematically prevents the separation of spatial coordinates, rendering an exact analytical solution impossible. To systematically resolve this challenge, we begin our theoretical treatment by applying Rayleigh-Schrödinger perturbation theory, isolating these problematic repulsion terms as a structural perturbation acting upon an exactly solvable independent-particle system.
\section{Perturbation Theory Approximation}
To systematically evaluate the electronic structure under a perturbative framework, the total non-relativistic Hamiltonian is partitioned into an unperturbed multi-particle base operator ($\hat{H}_0$) and a perturbation operator ($\hat{H}^{(1)}$) that accounts for the mutual inter-electronic repulsive interactions \cite[Sec. 9.1]{levine2014}:
\begin{equation}
\hat{H} = \hat{H}_0 + \hat{H}^{(1)}
\end{equation}
\subsection{Zeroth-Order Approximation}
In the zeroth-order approximation, the electron-electron Coulomb repulsion is entirely neglected. Consequently, the unperturbed Hamiltonian ($\hat{H}_0$) decomposes into a direct sum of independent, single-particle hydrogen-like operators \cite[Sec. 5.2]{griffiths2018}:
\begin{equation}
\hat{H}_0 = \hat{h}_1 + \hat{h}_2 + \hat{h}_3 = \sum_{i=1}^{3} \left[ -\frac{\hbar^2}{2m_e}\nabla_i^2 - \frac{Ze^2}{4\pi\varepsilon_0 r_i} \right]
\end{equation}
The corresponding unperturbed state satisfies the independent-particle Schrödinger equation \cite[Sec. 7.1]{griffiths2018}:
\begin{equation}
\hat{H}_0 \Psi^{(0)} = E^{(0)} \Psi^{(0)}
\end{equation}

\subsubsection{Zeroth-Order Wave Function Construction}
Because the unperturbed Hamiltonian contains no electronic cross-terms, the multi-particle spatial solution can be modeled as a simple product of independent, hydrogen-like single-particle orbitals \cite[Sec. 5.2]{griffiths2018}:
\begin{equation}
    \psi^{(0)}(\mathbf{r}_1, \mathbf{r}_2, \mathbf{r}_3) = \phi_{1s}(\mathbf{r}_1)\phi_{1s}(\mathbf{r}_2)\phi_{2s}(\mathbf{r}_3)
\end{equation}
These spatial functions are designated as \textit{hydrogen-like} because each electron effectively experiences an unshielded nuclear potential energy of $-Ze^2 / (4\pi\varepsilon_0 r_i)$ with $Z=3$, rendering the individual equations mathematically analogous to a single-electron system.

In general, any atomic spatial orbital $\phi$ can be decomposed into a product of a radial wave function ($R_{n,l}$) and a spherical harmonic ($Y_l^m$) representing the angular component \cite[Sec. 6.6]{levine2014}:
\begin{equation}
    \phi_{n,l,m}(\mathbf{r}) = R_{n,l}(r) Y_l^m(\theta, \varphi)
\end{equation}

For the ground-state configuration of lithium, the orbitals of interest are the $1s$ and $2s$ states. For $s$-orbitals (where the orbital angular momentum quantum number $l=0$), the wave function exhibits no angular dependence, and the spherical harmonic reduces to the constant $Y_0^0 = 1/\sqrt{4\pi}$. Multiplying this angular constant by the corresponding radial functions yields the following explicit spatial profiles evaluated in perturbation theory \cite[Sec. 6.6]{levine2014}:
\begin{equation}
    \phi_{1s}(\mathbf{r}) = \frac{1}{\sqrt{\pi}} \left( \frac{Z}{a_0} \right)^{3/2} e^{-Z r / a_0}
\end{equation}
\begin{equation}
    \phi_{2s}(\mathbf{r}) = \frac{1}{4\sqrt{2\pi}} \left( \frac{Z}{a_0} \right)^{3/2} \left( 2 - \frac{Z r}{a_0} \right) e^{-Z r / (2a_0)}
\end{equation}

To construct a complete, physically acceptable non-relativistic quantum state that satisfies anti-symmetrization, the spin degrees of freedom must be integrated to form individual spin-orbitals, defined as $u_k(\mathbf{x}_i) = \phi_k(\mathbf{r}_i)\sigma_k(s_i)$. Here, $\mathbf{x}_i = (\mathbf{r}_i, s_i)$ denotes the combined spatial and spin coordinates, and the spin functions are given by \cite[Sec. 4.4]{griffiths2018}:
\begin{equation}
    \sigma_k(s_i) = 
    \begin{cases} 
        \alpha(s_i), & \text{if } m_s = +\frac{1}{2} \\
        \beta(s_i), & \text{if } m_s = -\frac{1}{2} 
    \end{cases}
\end{equation}

By designating the spin-up and spin-down projection states as $\alpha$ and $\beta$ respectively, the active spin-orbitals for the $1s^2 2s^1$ ground-state configuration are formulated as \cite[Sec. 11.1]{levine2014}:
\begin{equation}
    u_1(\mathbf{x}_i) = \phi_{1s}(\mathbf{r}_i)\alpha(s_i), \quad u_2(\mathbf{x}_i) = \phi_{1s}(\mathbf{r}_i)\beta(s_i), \quad u_3(\mathbf{x}_i) = \phi_{2s}(\mathbf{r}_i)\alpha(s_i)
\end{equation}
According to the Pauli Exclusion Principle, no two indistinguishable fermions can occupy the same quantum state simultaneously. Therefore, the two electrons residing within the identical spatial $1s$ orbital must possess opposing spin projections ($m_s = +1/2$ and $m_s = -1/2$). The single valence electron occupying the $2s$ orbital can formally assume either spin projection without altering the scalar energy expectation values; hence, it is chosen arbitrarily as a spin-up ($\alpha$) state \cite[Sec. 5.1]{griffiths2018}.
In matrix representation, these spinors are explicitly defined as column vectors \cite[Sec. 4.4]{griffiths2018}:
\begin{equation}
    \alpha = \begin{pmatrix} 1 \\ 0 \end{pmatrix}, \quad \beta = \begin{pmatrix} 0 \\ 1 \end{pmatrix}
\end{equation}

For many-body perturbation integrations, these spin functions must satisfy the standard orthonormality relations. The normalization conditions for both projection states are expressed using formal spin-space integration as \cite[Sec. 10.1]{levine2014}:
\begin{equation}
    \int \alpha^*(s)\alpha(s)\,ds = 1, \quad \int \beta^*(s)\beta(s)\,ds = 1
\end{equation}
Furthermore, the orthogonality condition, which enforces the physical requirement that a single electron cannot simultaneously occupy opposing spin projections, is defined by:
\begin{equation}
    \int \alpha^*(s)\beta(s)\,ds = 0
\end{equation}

Because electrons are identical fermions, the overall multi-particle wave function must be totally antisymmetric under the permutation of any two particle indices ($P_{ij}\Psi^{(0)} = -\Psi^{(0)}$). This global symmetry requirement is rigorously satisfied by constructing a $3 \times 3$ Slater determinant framework ($\mathcal{M}$) \cite[Sec. 10.6]{levine2014}:
\begin{equation}
\Psi^{(0)} = \frac{1}{\sqrt{3!}} \det(\mathcal{M}) = \frac{1}{\sqrt{6}} \begin{vmatrix}
u_1(\mathbf{x}_1) & u_2(\mathbf{x}_1) & u_3(\mathbf{x}_1) \\
u_1(\mathbf{x}_2) & u_2(\mathbf{x}_2) & u_3(\mathbf{x}_2) \\
u_1(\mathbf{x}_3) & u_2(\mathbf{x}_3) & u_3(\mathbf{x}_3)
\end{vmatrix}
\end{equation}
where the arguments $(1, 2, 3)$ are explicitly written as $\mathbf{x}_i$ to represent the combined spatial and spin coordinates of the $i$-th electron. Expanding this determinant explicitly yields the complete linear combination consisting of six distinct permutation terms:
\begin{equation}
\begin{aligned}
\Psi^{(0)} = \frac{1}{\sqrt{6}} \Big[ & u_1(\mathbf{x}_1)u_2(\mathbf{x}_2)u_3(\mathbf{x}_3) - u_1(\mathbf{x}_1)u_3(\mathbf{x}_2)u_2(\mathbf{x}_3) \\
& + u_2(\mathbf{x}_1)u_3(\mathbf{x}_2)u_1(\mathbf{x}_3) - u_2(\mathbf{x}_1)u_1(\mathbf{x}_2)u_3(\mathbf{x}_3) \\
& + u_3(\mathbf{x}_1)u_1(\mathbf{x}_2)u_2(\mathbf{x}_3) - u_3(\mathbf{x}_1)u_2(\mathbf{x}_2)u_1(\mathbf{x}_3) \Big]
\end{aligned}
\end{equation}

\subsubsection{Systematic Pairwise Grouping}
To optimize mathematical efficiency and fully exploit orbital orthonormality during subsequent expectation value calculations, the six expanded permutation terms are clustered pairwise into three distinct symmetric brackets, denoted as $A$, $B$, and $C$:
\begin{equation}
\Psi^{(0)} = \frac{1}{\sqrt{6}} (A + B + C)
\end{equation}
where these sub-component functional groups are analytically partitioned as:
\begin{equation}
\begin{aligned}
A &= u_1(\mathbf{x}_1)u_2(\mathbf{x}_2)u_3(\mathbf{x}_3) - u_2(\mathbf{x}_1)u_1(\mathbf{x}_2)u_3(\mathbf{x}_3) = \big[u_1(\mathbf{x}_1)u_2(\mathbf{x}_2) - u_2(\mathbf{x}_1)u_1(\mathbf{x}_2)\big] u_3(\mathbf{x}_3) \\
B &= u_3(\mathbf{x}_1)u_1(\mathbf{x}_2)u_3(\mathbf{x}_3) - u_1(\mathbf{x}_1)u_3(\mathbf{x}_2)u_2(\mathbf{x}_3) = \big[u_3(\mathbf{x}_1)u_1(\mathbf{x}_2) - u_1(\mathbf{x}_1)u_3(\mathbf{x}_2)\big] u_2(\mathbf{x}_3) \\
C &= u_2(\mathbf{x}_1)u_3(\mathbf{x}_2)u_1(\mathbf{x}_3) - u_3(\mathbf{x}_1)u_2(\mathbf{x}_2)u_1(\mathbf{x}_3) = \big[u_2(\mathbf{x}_1)u_3(\mathbf{x}_2) - u_3(\mathbf{x}_1)u_2(\mathbf{x}_2)\big] u_1(\mathbf{x}_3)
\end{aligned}
\end{equation}

\subsubsection{Zeroth-Order Energy Baseline}
The unperturbed single-particle energy eigenvalues for a hydrogen-like system are determined via the analytical solution to the radial Schrödinger equation \cite[Sec. 6.6]{levine2014}:
\begin{equation}
E_n = -\left( \frac{m_e e^4}{2(4\pi\varepsilon_0)^2 \hbar^2} \right) \frac{Z^2}{n^2} \approx -13.6057\,\text{eV} \cdot \frac{Z^2}{n^2}
\end{equation}

For the lithium nucleus ($Z=3$), the constituent physical constants are defined as follows: the electron rest mass $m_e = 0.5109989\,\mathrm{MeV}/c^2$, the elementary charge $e = 1.602176 \times 10^{-19}\,\mathrm{C}$, the vacuum permittivity $\varepsilon_0 = 8.854188 \times 10^{-12}\,\mathrm{F/m}$, and the reduced Planck constant $\hbar = 6.582120 \times 10^{-16}\,\mathrm{eV\cdot s}$. The pre-factor encapsulates the fundamental Rydberg energy unit ($-13.6057\,\text{eV}$), yielding the individual unshielded orbital energy levels directly as:
\begin{equation}
E_{1s} = -13.6057 \left( \frac{3^2}{1^2} \right) \approx -122.45\,\text{eV}
\end{equation}
\begin{equation}
E_{2s} = -13.6057 \left( \frac{3^2}{2^2} \right) \approx -30.61\,\text{eV}
\end{equation}

Since the unperturbed total Hamiltonian operator is a direct linear sum of non-interacting components, the total zeroth-order energy baseline is determined by the cumulative energies of the occupied orbital states ($2E_{1s} + E_{2s}$):
\begin{equation}
E^{(0)} = 2(-122.45\,\text{eV}) + (-30.61\,\text{eV}) = -275.51\,\text{eV}
\end{equation}
\subsection{First-Order Perturbation Theory}
The perturbation Hamiltonian accounting for the cumulative inter-electronic Coulomb repulsion is expressed as \cite[Sec. 11.7]{levine2014}:
\begin{equation}
\hat{H}^{(1)} = \hat{H}_{12}^{(1)} + \hat{H}_{13}^{(1)} + \hat{H}_{23}^{(1)} = \frac{e^2}{4\pi\varepsilon_0 r_{12}} + \frac{e^2}{4\pi\varepsilon_0 r_{13}} + \frac{e^2}{4\pi\varepsilon_0 r_{23}}
\end{equation}
where the indices represent the pairwise interactions formed through the circular permutation of the three electrons, with $r_{ij} = |\mathbf{r}_i - \mathbf{r}_j|$. The corresponding first-order energy correction is evaluated as the expectation value of this operator over the unperturbed state \cite[Sec. 7.1]{griffiths2018}:
\begin{equation}
E^{(1)} = \langle \Psi^{(0)} | \hat{H}^{(1)} | \Psi^{(0)} \rangle
\end{equation}

Since the unperturbed wave function $\Psi^{(0)}$ is completely antisymmetric under coordinate exchange and the individual electrons are fundamentally indistinguishable, the inner product over each pairwise repulsion operator yields identical mathematical results \cite[Sec. 10.5]{levine2014}:
\begin{equation}
\langle \Psi^{(0)} | \hat{H}_{12}^{(1)} | \Psi^{(0)} \rangle = \langle \Psi^{(0)} | \hat{H}_{13}^{(1)} | \Psi^{(0)} \rangle = \langle \Psi^{(0)} | \hat{H}_{23}^{(1)} | \Psi^{(0)} \rangle
\end{equation}

This permutational symmetry simplifies the total expectation value down to a single interaction channel scaled by a combinatorial factor of three:
\begin{equation}
E^{(1)} = 3 \langle \Psi^{(0)} | \hat{H}_{12}^{(1)} | \Psi^{(0)} \rangle
\end{equation}

Expanding this expression using the previously established pairwise grouped brackets $A$, $B$, and $C$ (Equation 18) yields:
\begin{equation}
\begin{aligned}
E^{(1)} &= 3 \cdot \left( \frac{1}{6} \right) \langle A + B + C | \hat{H}_{12}^{(1)} | A + B + C \rangle \\
&= \frac{1}{2} \Big[ \langle A | \hat{H}_{12}^{(1)} | A \rangle + \langle B | \hat{H}_{12}^{(1)} | B \rangle + \langle C | \hat{H}_{12}^{(1)} | C \rangle \\
&\quad + \langle A | \hat{H}_{12}^{(1)} | B \rangle + \langle B | \hat{H}_{12}^{(1)} | A \rangle + \langle A | \hat{H}_{12}^{(1)} | C \rangle \\
&\quad + \langle C | \hat{H}_{12}^{(1)} | A \rangle + \langle B | \hat{H}_{12}^{(1)} | C \rangle + \langle C | \hat{H}_{12}^{(1)} | B \rangle \Big]
\end{aligned}
\end{equation}

\subsubsection{Orthogonality Reduction to Coulomb and Exchange Integrals}
The matrix elements within Eq.(28) can be drastically simplified by integrating over the coordinates of the non-interacting unperturbed spectator electron. Leveraging the spatial and spin orthonormality of the single-particle spin-orbitals ($\langle u_i | u_j \rangle = \delta_{ij}$) \cite[Sec. 2.2]{griffiths2018}, only the three diagonal matrix elements ($AA$, $BB$, and $CC$) provide non-vanishing contributions, which are evaluated as follows:

\textbf{i. Evaluation of $\langle A | \hat{H}_{12}^{(1)} | A \rangle$:} In this configuration, electron 3 acts as the spectator particle residing in state $u_3$. Utilizing the normalization condition $\langle u_3(\mathbf{x}_3)|u_3(\mathbf{x}_3) \rangle = 1$, the integration reduces to the spatial and spin coordinates of the two active interacting electrons ($\mathbf{x}_1$ and $\mathbf{x}_2$) \cite[Sec. 10.7]{levine2014}:
\begin{equation}
\begin{aligned}
\langle A|\hat{H}_{12}^{(1)}|A \rangle &= \iint [u_1(\mathbf{x}_1)u_2(\mathbf{x}_2) - u_2(\mathbf{x}_1)u_1(\mathbf{x}_2)]^* \hat{H}_{12}^{(1)} [u_1(\mathbf{x}_1)u_2(\mathbf{x}_2) - u_2(\mathbf{x}_1)u_1(\mathbf{x}_2)] \,d\mathbf{x}_1 \,d\mathbf{x}_2 \\
&= 2 \iint |\phi_{1s}(\mathbf{r}_1)|^2 \frac{e^2}{4\pi\varepsilon_0 |\mathbf{r}_1 - \mathbf{r}_2|} |\phi_{1s}(\mathbf{r}_2)|^2 \,d^3\mathbf{r}_1 \,d^3\mathbf{r}_2 \\
&= 2J_{1s,1s}
\end{aligned}
\end{equation}
where $J_{1s,1s}$ denotes the direct Coulomb integral for the two interacting electrons within the $1s$ core shell.

\textbf{ii. Evaluation of $\langle B | \hat{H}_{12}^{(1)} | B \rangle$:} Here, electron 3 occupies the spectator state $u_2$. Performing the integration over the remaining active spin-orbital coordinates yields \cite[Sec. 10.7]{levine2014}:
\begin{equation}
\begin{aligned}
\langle B|\hat{H}_{12}^{(1)}|B \rangle &= \iint [u_3(\mathbf{x}_1)u_1(\mathbf{x}_2) - u_1(\mathbf{x}_1)u_3(\mathbf{x}_2)]^* \hat{H}_{12}^{(1)} [u_3(\mathbf{x}_1)u_1(\mathbf{x}_2) - u_1(\mathbf{x}_1)u_3(\mathbf{x}_2)] \,d\mathbf{x}_1 \,d\mathbf{x}_2 \\
&= 2 \iint |\phi_{1s}(\mathbf{r}_1)|^2 \frac{e^2}{4\pi\varepsilon_0|\mathbf{r}_1 - \mathbf{r}_2|} |\phi_{2s}(\mathbf{r}_2)|^2 \,d^3\mathbf{r}_1 \,d^3\mathbf{r}_2 \\
&\quad - 2 \iint \phi_{1s}^*(\mathbf{r}_1) \phi_{2s}^*(\mathbf{r}_2) \frac{e^2}{4\pi\varepsilon_0|\mathbf{r}_1 - \mathbf{r}_2|} \phi_{1s}(\mathbf{r}_2) \phi_{2s}(\mathbf{r}_1) \,d^3\mathbf{r}_1 \,d^3\mathbf{r}_2 \\
&= 2J_{1s,2s} - 2K_{1s,2s}
\end{aligned}
\end{equation}
where $J_{1s,2s}$ and $K_{1s,2s}$ represent the direct Coulomb and quantum exchange integrals, respectively, between the $1s$ core and $2s$ valence states.

\textbf{iii. Evaluation of $\langle C | \hat{H}_{12}^{(1)} | C \rangle$:} In this bracket, electron 3 is the spectator particle in state $u_1$. Because $\langle u_1(\mathbf{x}_3)|u_1(\mathbf{x}_3) \rangle = 1$, the matrix element evaluates to \cite[Sec. 10.7]{levine2014}:
\begin{equation}
\begin{aligned}
\langle C|\hat{H}_{12}^{(1)}|C \rangle &= \iint [u_2(\mathbf{x}_1)u_3(\mathbf{x}_2) - u_3(\mathbf{x}_1)u_2(\mathbf{x}_2)]^* \hat{H}_{12}^{(1)} [u_2(\mathbf{x}_1)u_3(\mathbf{x}_2) - u_3(\mathbf{x}_1)u_2(\mathbf{x}_2)] \,d\mathbf{x}_1 \,d\mathbf{x}_2 \\
&= 2 \iint |\phi_{1s}(\mathbf{r}_1)|^2 \frac{e^2}{4\pi\varepsilon_0|\mathbf{r}_1 - \mathbf{r}_2|} |\phi_{2s}(\mathbf{r}_2)|^2 \,d^3\mathbf{r}_1 \,d^3\mathbf{r}_2 \\
&= 2J_{1s,2s}
\end{aligned}
\end{equation}
Notably, the expected exchange term in the $C$-bracket vanishes due to explicit spin orthogonality, as the corresponding spin integral contains the factors $\langle \alpha|\beta\rangle = 0$ and $\langle \beta|\alpha\rangle = 0$.

All cross-terms (such as $\langle A | \hat{H}_{12}^{(1)} | B \rangle$) vanish entirely because the spectator electron maps to mutually orthogonal spin-orbitals (e.g., $\langle u_3 | u_2 \rangle = 0$ or $\langle u_3 | u_1 \rangle = 0$). Collecting the surviving non-zero components and substituting them back into the primary expression yields:
\begin{equation}
E^{(1)} = \frac{1}{2} \Big[ 2J_{1s,1s} + \big(2J_{1s,2s} - 2K_{1s,2s}\big) + 2J_{1s,2s} \Big] = J_{1s,1s} + 2J_{1s,2s} - K_{1s,2s}
\end{equation}

\subsubsection{Numerical Substitution and First-Order Energy Results}
The spatial double integrals defining these static Coulomb and exchange interactions are explicitly written as \cite[Sec. 10.5]{levine2014}:
\begin{equation}
J_{1s,1s} = \iint \phi_{1s}^*(\mathbf{r}_1)\phi_{1s}^*(\mathbf{r}_2) \left(\frac{e^2}{4\pi\varepsilon_0 r_{12}}\right) \phi_{1s}(\mathbf{r}_1)\phi_{1s}(\mathbf{r}_2) \,d^3\mathbf{r}_1 \,d^3\mathbf{r}_2
\end{equation}
\begin{equation}
J_{1s,2s} = \iint \phi_{1s}^*(\mathbf{r}_1)\phi_{2s}^*(\mathbf{r}_2) \left(\frac{e^2}{4\pi\varepsilon_0 r_{12}}\right) \phi_{1s}(\mathbf{r}_1)\phi_{2s}(\mathbf{r}_2) \,d^3\mathbf{r}_1 \,d^3\mathbf{r}_2
\end{equation}
\begin{equation}
K_{1s,2s} = \iint \phi_{1s}^*(\mathbf{r}_1)\phi_{2s}^*(\mathbf{r}_2) \left(\frac{e^2}{4\pi\varepsilon_0 r_{12}}\right) \phi_{2s}(\mathbf{r}_1)\phi_{1s}(\mathbf{r}_2) \,d^3\mathbf{r}_1 \,d^3\mathbf{r}_2
\end{equation}

The exact analytical evaluations of these multi-center integrals for the lithium nucleus ($Z=3$) yield the following expressions (the detailed integral derivations are provided in Appendix A):
\begin{equation}
    J_{1s,1s} = \frac{5}{8} \left(\frac{Z e^2}{4\pi\varepsilon_0 a_0}\right), \quad 
    J_{1s,2s} = \frac{17}{81} \left(\frac{Z e^2}{4\pi\varepsilon_0 a_0}\right), \quad 
    K_{1s,2s} = \frac{16}{729} \left(\frac{Z e^2}{4\pi\varepsilon_0 a_0}\right)
\end{equation}

Factoring out the core unit energy group $\frac{e^2}{4\pi\varepsilon_0 a_0} = 2\,\text{Ry} \approx 27.2114\,\text{eV}$, the total first-order energy correction for $Z=3$ evaluates quantitatively to:
\begin{equation}
    E^{(1)} = 3\left[\left( \frac{5}{8} \right) + 2\left(\frac{17}{81}\right) - \left( \frac{16}{729} \right)\right] \times 27.2114\,\text{eV} \approx 83.50\,\text{eV}
\end{equation}
which is in precise agreement with standard literature references \cite[Sec. 10.7]{levine2014}. 

Consequently, the total cumulative ground-state energy corrected to first order is given by:
\begin{equation}
E_{\text{total}}^{(1)} = E^{(0)} + E^{(1)} \approx -275.51\,\text{eV} + 83.50\,\text{eV} = -192.01\,\text{eV}
\end{equation}
While this first-order correction significantly resolves the unphysical baseline energy, it still deviates from the reference non-relativistic reference ground-state energy of lithium ($\sim -203.5\ \text{eV}$) by approximately $5.65\%$. This residual error underscores the necessity of computing second-order perturbation corrections to account for dynamic electron correlation effects.
The reference value is adopted from \cite{nist2024} and the total ionization energy is calculated via adding up the three values ($ -5.391714996\ \text{eV} -75.6400970\ \text{eV} -122.45435913\ \text{eV} = -203.486119\,\text{eV}$) shown in the reference, where the first ionization energy is measured experimentally, whereas the other two values are calculated theoretically. The final result is rounded herein $-203.5\ \text{eV}$.
\subsection{Second-Order Perturbation Theory}
While first-order perturbation theory evaluates electron-electron repulsion strictly within an averaged-field framework, incorporating dynamical, instantaneous electronic correlation and virtual state mixing necessitates higher-order corrections \cite[16.3]{levine2014} \cite[Chap. 6]{szabo1982}. The second-order energy correction can be formally derived from the time-independent Schrödinger equation $\hat{H}|\Psi\rangle = E|\Psi\rangle$ by expanding the operators and states in terms of a continuous ordering parameter $\lambda$ \cite[Sec. 9.2]{levine2014} \cite{kalhous2004}:
\begin{align}
    \hat{H} &= \hat{H}_0 + \lambda \hat{H}^{(1)} \nonumber \\
    |\Psi\rangle &= |\Psi_0^{(0)}\rangle + \lambda |\Psi^{(1)}\rangle + \lambda^2 |\Psi^{(2)}\rangle + \dots \nonumber \\
    E &= E_0^{(0)} + \lambda E^{(1)} + \lambda^2 E^{(2)} + \dots
\end{align}

Substituting these series expansions into the Schrödinger equation and isolating the second-order ($\lambda^2$) components establishes the fundamental relation \cite[Sec. 7.1]{griffiths2018}:
\begin{equation}
    \hat{H}_0 |\Psi^{(2)}\rangle + \hat{H}^{(1)} |\Psi^{(1)}\rangle = E_0^{(0)} |\Psi^{(2)}\rangle + E^{(1)} |\Psi^{(1)}\rangle + E^{(2)} |\Psi_0^{(0)}\rangle
\end{equation}

Projecting this relation onto the unperturbed ground-state bra $\langle\Psi_0^{(0)}|$ and utilizing the Hermiticity of the unperturbed Hamiltonian ($\langle\Psi_0^{(0)}|\hat{H}_0 = E_0^{(0)} \langle\Psi_0^{(0)}|$), the unknown second-order wave function components cancel out identically. Assuming an intermediate-normalized zeroth-order ground state ($\langle\Psi_0^{(0)}|\Psi_0^{(0)}\rangle = 1$), the expression reduces to the compact bracket formulation \cite[Sec. 7.1]{griffiths2018}:
\begin{equation}
    E^{(2)} = \langle\Psi_0^{(0)}| \hat{H}^{(1)} - E^{(1)} |\Psi^{(1)}\rangle
\end{equation}

To resolve this expression into an explicitly calculable sum over static unperturbed states, the first-order wave function correction $|\Psi^{(1)}\rangle$—which represents the explicit multi-body polarization of the doublet state under electron correlation—is expanded as a linear combination of the unperturbed excited eigenstates $|\Psi_m^{(0)}\rangle$ \cite[Sec. 7.1]{griffiths2018}:
\begin{equation}
    |\Psi^{(1)}\rangle = \sum_{m\ne0} c_m |\Psi_{m}^{(0)}\rangle \quad \text{where} \quad c_m = \frac{\langle\Psi_{m}^{(0)}|\hat{H}^{(1)}|\Psi_{0}^{(0)}\rangle}{E_{0}^{(0)} - E_{m}^{(0)}}
\end{equation}

Substituting this expansion back into the bracket relation yields the standard Rayleigh-Schrödinger second-order energy correction formula:
\begin{equation}
    E^{(2)} = \sum_{m \neq 0} \frac{\left| \langle \Psi_m^{(0)} | \hat{H}^{(1)} | \Psi_0^{(0)} \rangle \right|^2}{E_0^{(0)} - E_m^{(0)}}
\end{equation}

At this stage, it is crucial to address the intrinsic spin degeneracy of the ground state. The unperturbed lithium configuration ($1s^2 2s^1$) is two-fold degenerate because the valence electron in the $2s$ orbital can assume either a spin-up ($m_s = +1/2$) or a spin-down ($m_s = -1/2$) projection, both yielding the identical unperturbed energy $E_0^{(0)}$. The presence of degeneracy formally requires the framework of Degenerate Perturbation Theory (DPT), which dictates the construction and diagonalization of the perturbation matrix within this degenerate subspace \cite[Sec. 10.4]{levine2014}:
\begin{equation}
    \mathbf{H}^{(1)} = \begin{pmatrix}
    \langle \alpha | \hat{H}^{(1)} | \alpha \rangle & \langle \alpha | \hat{H}^{(1)} | \beta \rangle \\
    \langle \beta | \hat{H}^{(1)} | \alpha \rangle & \langle \beta | \hat{H}^{(1)} | \beta \rangle
    \end{pmatrix}
\end{equation}

However, the physical perturbation operator—the inter-electronic Coulomb repulsion $\hat{H}^{(1)} = \sum \frac{e^2}{4\pi\varepsilon_0 r_{ij}}$—is purely spatial and contains no spin-dependent components. Because this operator commutes with the total spin operators, the quantum mechanical integrals separate into independent spatial and spin products. For the off-diagonal elements [in eq. 44], the inner product of the spin states strictly enforces the spin orthogonality condition:
\begin{equation}
    \langle \alpha | \hat{H}^{(1)} | \beta \rangle \propto \int \alpha^*(s) \beta(s) \,ds = 0
\end{equation}

Consequently, the cross-subspace coupling terms vanish entirely, demonstrating that the perturbation matrix is inherently diagonal. Furthermore, since the spatial integrals for the diagonal elements are physically identical, no lifting of the degeneracy occurs. Because the perturbation does not couple or split these degenerate states, the DPT matrix equation mathematically collapses directly into the standard Non-Degenerate Perturbation Theory (NDPT) framework \cite[Sec. 10.4]{levine2014}. This spatial symmetry justifies the direct application of the non-degenerate second-order summation formula across individual orbital channels.

\subsubsection{Excited Slater Matrix and Integral Reduction}
To evaluate the matrix elements for the excited states, we insert an unpopulated higher virtual single-particle orbital, such as $u_4 = \phi_{3s}\alpha$, into the system configuration space to construct the excited Slater matrix ($\mathcal{M}_m$):
\begin{equation}
\Psi_m^{(0)} = \frac{1}{\sqrt{6}} \det(\mathcal{M}_m) = \frac{1}{\sqrt{6}} \begin{vmatrix}
u_1(\mathbf{x}_1) & u_2(\mathbf{x}_1) & u_4(\mathbf{x}_1) \\
u_1(\mathbf{x}_2) & u_2(\mathbf{x}_2) & u_4(\mathbf{x}_2) \\
u_1(\mathbf{x}_3) & u_2(\mathbf{x}_3) & u_4(\mathbf{x}_3)
\end{vmatrix}
\end{equation}

Expanding this determinant explicitly along the first row yields the linear combination of six distinct permutation terms:
\begin{equation}
\begin{aligned}
    \Psi_{m}^{(0)} = \frac{1}{\sqrt{6}} \Big[ &u_1(\mathbf{x}_1)u_2(\mathbf{x}_2)u_4(\mathbf{x}_3) - u_1(\mathbf{x}_1)u_4(\mathbf{x}_2)u_2(\mathbf{x}_3) \\
    &- u_2(\mathbf{x}_1)u_1(\mathbf{x}_2)u_4(\mathbf{x}_3) + u_2(\mathbf{x}_1)u_4(\mathbf{x}_2)u_1(\mathbf{x}_3) \\
    &+ u_4(\mathbf{x}_1)u_1(\mathbf{x}_2)u_2(\mathbf{x}_3) - u_4(\mathbf{x}_1)u_2(\mathbf{x}_2)u_1(\mathbf{x}_3) \Big]
\end{aligned}
\end{equation}

By systematically rearranging these terms, the active spatial and spin pairs can be isolated. These six expanded permutation terms are clustered pairwise into three distinct grouped brackets denoted as $A_m$, $B_m$, and $C_m$:
\begin{equation}
\Psi_m^{(0)} = \frac{1}{\sqrt{6}} (A_m + B_m + C_m)
\end{equation}
where the excited multi-particle sub-components are defined as:
\begin{equation}
\begin{aligned}
A_m &= \big[u_1(\mathbf{x}_1)u_2(\mathbf{x}_2) - u_2(\mathbf{x}_1)u_1(\mathbf{x}_2)\big] u_4(\mathbf{x}_3) \\
B_m &= \big[u_4(\mathbf{x}_1)u_1(\mathbf{x}_2) - u_1(\mathbf{x}_1)u_4(\mathbf{x}_2)\big] u_2(\mathbf{x}_3) \\
C_m &= \big[u_2(\mathbf{x}_1)u_4(\mathbf{x}_2) - u_4(\mathbf{x}_1)u_2(\mathbf{x}_2)\big] u_1(\mathbf{x}_3)
\end{aligned}
\end{equation}

Projecting the pairwise perturbation operator $\hat{H}_{12}^{(1)}$ across these excited configurations simplifies the multi-body channels directly into specific spatial transition integrals over the joint spatial-spin coordinates ($d\mathbf{x}_1 \,d\mathbf{x}_2$):
\begin{equation}
\begin{gathered}
    \langle A_m | \hat{H}_{12}^{(1)} | A \rangle = 0 \\[0.3cm]
    \begin{aligned}
        \langle B_m | \hat{H}_{12}^{(1)} | B \rangle &= \iint [u_4(\mathbf{x}_1)u_1(\mathbf{x}_2) - u_1(\mathbf{x}_1)u_4(\mathbf{x}_2)]^* \hat{H}_{12}^{(1)} [u_3(\mathbf{x}_1)u_1(\mathbf{x}_2) - u_1(\mathbf{x}_1)u_3(\mathbf{x}_2)] \,d\mathbf{x}_1 \,d\mathbf{x}_2 \\
        &= 2J_{3s,1s}^{\prime} - 2K_{3s,1s}^{\prime} \\[0.3cm]
        \langle C_m | \hat{H}_{12}^{(1)} | C \rangle &= \iint [u_2(\mathbf{x}_1)u_4(\mathbf{x}_2) - u_4(\mathbf{x}_1)u_2(\mathbf{x}_2)]^* \hat{H}_{12}^{(1)} [u_2(\mathbf{x}_1)u_3(\mathbf{x}_2) - u_3(\mathbf{x}_1)u_2(\mathbf{x}_2)] \,d\mathbf{x}_1 \,d\mathbf{x}_2 \\
        &= 2J_{3s,1s}^{\prime}
    \end{aligned}
\end{gathered}
\end{equation}

Consequently, the total aggregated transition matrix element ($M$) reduces to the following linear combination:
\begin{equation}
M = \langle \Psi_m^{(0)} | \hat{H}_{12}^{(1)} | \Psi_0^{(0)} \rangle = 2J'_{3s,1s} - K'_{3s,1s}
\end{equation}

The defining spatial double integrals for these virtual transition states are given by \cite[Sec. 7.4]{bransden1983}:
\begin{equation}
J'_{3s,1s} = \iint \phi_{3s}^*(\mathbf{r}_1)\phi_{1s}^*(\mathbf{r}_2) \left(\frac{e^2}{4\pi\varepsilon_0 r_{12}}\right) \phi_{2s}(\mathbf{r}_1)\phi_{1s}(\mathbf{r}_2) \,d^3\mathbf{r}_1 \,d^3\mathbf{r}_2
\end{equation}
\begin{equation}
K'_{3s,1s} = \iint \phi_{3s}^*(\mathbf{r}_1)\phi_{1s}^*(\mathbf{r}_2) \left(\frac{e^2}{4\pi\varepsilon_0 r_{12}}\right) \phi_{1s}(\mathbf{r}_1)\phi_{2s}(\mathbf{r}_2) \,d^3\mathbf{r}_1 \,d^3\mathbf{r}_2
\end{equation}

To evaluate the energy denominator [in eq. 43], the unperturbed zeroth-order energy of the excited $1s^2 3s^1$ configuration ($E_{3s}^{(0)}$) must be determined. Within the independent-particle model, the energy of a hydrogen-like state is dictated by:
\begin{equation}
    E_n = -13.6057\,\text{eV} \cdot \frac{Z^2}{n^2}
\end{equation}

For the lithium atom ($Z=3$), the individual single-particle energy eigenvalues for the $1s$ and $3s$ states evaluate to:
\begin{align}
    E_{1s} &= -13.6057 \cdot \frac{3^2}{1^2} \approx -122.45\,\text{eV} \\
    E_{3s} &= -13.6057 \cdot \frac{3^2}{3^2} \approx -13.61\,\text{eV}
\end{align}

Since the excited configuration consists of two core electrons in the $1s$ orbital and a single valence electron promoted to the $3s$ orbital, the total unperturbed energy is the linear sum of these occupied states:
\begin{equation}
    E_{3s}^{(0)} = 2E_{1s} + E_{3s} \approx 2(-122.45\,\text{eV}) + (-13.61\,\text{eV}) = -258.51\,\text{eV}
\end{equation}

The resulting energy denominator required for the second-order perturbation expansion is given by:
\begin{equation}
    E_0^{(0)} - E_{3s}^{(0)} \approx -275.51\,\text{eV} - (-258.51\,\text{eV}) = -17.00\,\text{eV}
\end{equation}
Substituting the evaluated transition matrix element and the energy denominator into the second-order expansion yields the specific single-excitation correction:
\begin{equation}
    E_{3s}^{(2)} = \frac{(2J'_{3s,1s} - K'_{3s,1s})^2}{E_0^{(0)} - E_{3s}^{(0)}}
\end{equation}

The analytical numerical integration of these spatial double transition integrals yields the following values:
\begin{equation}
    J'_{3s,1s} \approx 4.12\,\text{eV}, \quad K'_{3s,1s} \approx 0.92\,\text{eV}
\end{equation}
Substituting these transition values into the second-order energy expression results quantitatively in:
\begin{equation}
E_{3s}^{(2)} = \frac{(2(4.12) - 0.92)^2}{-17.00} \approx -3.16\,\text{eV}
\end{equation}

As demonstrated, the analytical evaluation of virtual state mixings involves highly intricate multi-center integrations that are practically restricted to spherically symmetric channels such as the $3s$ state. While physical electron correlation is heavily driven by excitations into higher angular momentum states (such as $p$ and $d$ channels), individual single-electron transitions to these states are strictly forbidden by selection rules due to the spherical symmetry ($L=0$) of the ground state and the scalar nature of the Coulomb operator. Consequently, the remaining accessible single-excitation channels are confined strictly to higher $s$-orbitals (e.g., $4s$, $5s$, and $6s$).

To capture these successive single-particle correlation effects without encountering analytical intractability, the numerical integrations for higher virtual transitions were evaluated computationally using an optimized Python script. The resulting second-order energy corrections for these discrete excited configurations are tabulated in Appendix B, alongside comprehensive computational methodologies.

Summing the unperturbed reference baseline and the cumulative perturbation corrections—where the total second-order correction $E^{(2)}$ converges numerically to approximately $-4.35\,\text{eV}$—yields the total energy of the system:
\begin{equation}
E_{\text{total}} = E^{(0)} + E^{(1)} + E^{(2)}
\end{equation}
\begin{equation}
E_{\text{total}} \approx -275.51\,\text{eV} + 83.50\,\text{eV} - 4.35\,\text{eV} = -196.36\,\text{eV}
\end{equation}

\subsubsection{Virtual Double-Electron Excitations}
Although single-electron virtual transitions are constrained by selection rules exclusively to higher $s$-orbitals, a significant portion of the residual correlation energy is governed by simultaneous double-electron excitations, where the conservation of total angular momentum is preserved globally without requiring individual orbital constraints. This missing energy component—excluding minor relativistic contributions—originates predominantly from coupled two-electron excitations involving the valence $2s$ electron and one of the core $1s$ electrons. Evaluating these double excitations demands mapping out non-spherical configurations ($p^2, d^2, f^2, g^2$) and isolating combined multi-particle states that satisfy a total angular momentum of $L=0$.

To preserve physical validity, the total angular momentum of the excited two-electron subsystem must couple exclusively to a total value of 0 or 1, thereby conserving the total atomic angular momentum at $L=0$. These coupling constraints dictate that when the unexcited spectator core electron is in a spin-up state ($\alpha$), the excited subsystem must couple to a singlet state ($S=0$), whereas a spin-down spectator ($\beta$) necessitates a triplet coupling ($S=1$). According to rigorous configurations maps established in literature \cite{weiss1961}, the total number of valid multi-particle configurations satisfying these conditions is forty-five for the lithium atom. The multi-body wave function encompassing these expansions can be structured as:
\begin{equation}
    \Psi_{\text{3E}} = K \cdot 2s^{\prime\prime} + \Phi_{1} + \Phi_{2}
\end{equation}
where $2s''$ represents the Slater determinant adjusted for the coupled electrons, while $\Phi_1$ and $\Phi_2$ represent the multi-body interactions between the outer valence electron and the inner core electron of opposite and parallel spins, respectively, defined explicitly as:
\begin{align}
\Phi_{1} &= (1s)^{2}1s^{\prime\prime} + (1s1s^{\prime})1s^{\prime\prime} + (2p)^{2}1s^{\prime\prime} + (1s)^{2}2s + (2p^{\prime\prime})^{2}1s + (3d^{\prime})^{2}1s \\
\Phi_{2} &= (3p2p^{\prime\prime})[^3S]1s + (2p^{\prime\prime}3p^{\prime\prime})[^3S]1s + (5d3d^{\prime})[^3S]1s + (2p^{\prime\prime}3p^{\prime\prime})[^3S]2s
\end{align}

The listed configurations are the subset of excited-electron arrangements that satisfy the required quantum numbers of the target Lithium state,where the orbital symbols ($1s$ $2p$ $3d$ etc.) specify the occupied orbital symmetries, primes (both $'$ and $''$ ) simply do label different correlated orbitals of the same type as the unprimed ones and do not refer to "more excitement" etc., and the symbol $^{3}S$ after the spatial configurations of some of the pairs denotes that that pair is specifically coupled to a triplet S state. Incorporating these simultaneous double excitations up to the virtual $5g$ shell yields an additional correlation correction of $-0.814\,\text{eV}$ \cite{weiss1961}. This shifts the total converged non-relativistic energy to $-197.174\,\text{eV}$, successfully reducing the relative error against the reference baseline \cite{nist2024} to $3.11\%$. This result is integrated into the scope of this work to demonstrate the physical significance of simultaneous virtual multi-particle excitations and to expand the theoretical framework. Because these terms were extracted from reference data rather than evaluated directly, this corrected value is omitted from subsequent graphical comparisons and discussion sections.

\subsubsection{Comparison of Zeroth, First, and Second-Order Energy Corrections}
\begin{figure}[h]
       \centering
       \includegraphics[width=1\linewidth]{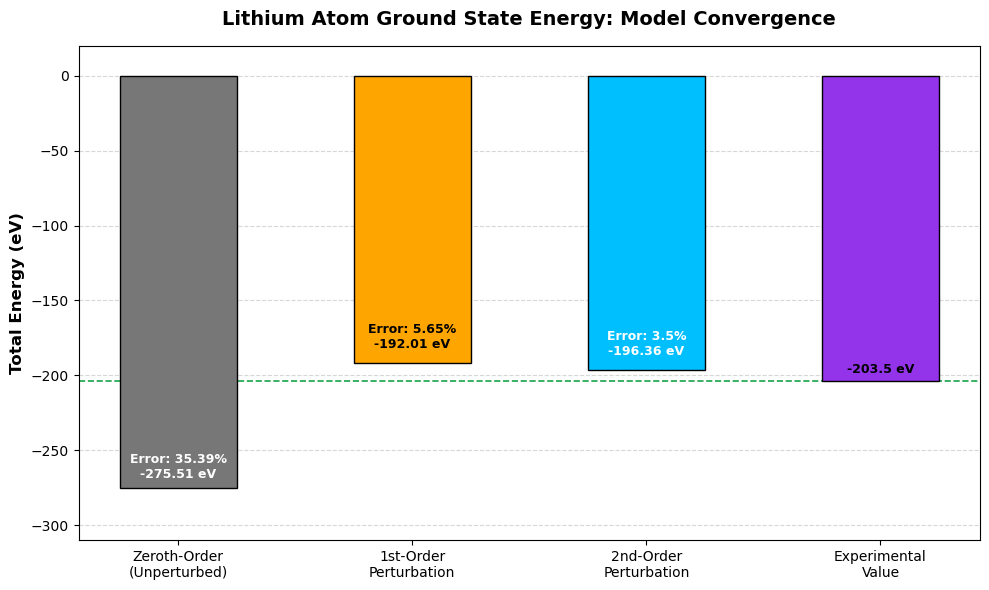}

      \caption{Total ground-state energy convergence of the lithium atom across successive perturbation orders compared against the non-relativistic reference baseline. The percentage metrics indicate the relative error calculated with respect to the reference value ($-203.5\,\text{eV}$) \cite{nist2024}.}
    \label{fig:1}
\end{figure}

A systematic comparison of the sequential perturbation levels reveals a monotonic convergence toward the non-relativistic reference ground-state energy of approximately $-203.5\,\text{eV}$ \cite{nist2024}. As illustrated in Fig.~\ref{fig:1}, the zeroth-order approximation, which entirely neglects inter-electronic repulsion, yields a severely underestimated energy of $-275.51\,\text{eV}$, resulting in a substantial relative error of $35.84\%$. Introducing the static Coulomb and exchange interactions via the first-order correction drastically shifts the energy eigenvalue upward to $-192.01\,\text{eV}$, minimizing the error to $5.65\%$. Finally, incorporating dynamic electron correlation through second-order virtual $s$-orbital transitions refines the total calculated energy to $-196.36\,\text{eV}$, driving the relative error down to $3.5\%$. 

This systematic trend quantitatively demonstrates how consecutive higher-order perturbation terms successfully diminish physical errors by capturing the missing electron correlation energy. Physically, the unperturbed baseline is unrealistically negative because the non-interacting model overestimates the binding energy between the unshielded nucleus and the individual electrons. The subsequent first and second-order corrections introduce the necessary repulsive potential energy, shifting the eigenvalues upward and systematically converging toward the exact reference benchmark.

\
\vspace{5mm}

\subsubsection{Higher-Order Perturbation Corrections}
Although first and second-order perturbation theories capture the bulk of static Coulomb repulsion and primary pair-correlation effects, achieving full spectroscopic accuracy necessitates the consideration of higher-order terms ($n \geq 3$). These terms account for increasingly complex multi-electronic virtual excitations and non-linear correlation couplings. Within the Rayleigh-Schrödinger framework under intermediate normalization ($\langle \Psi_0^{(0)} | \Psi_0^{(k)} \rangle = \delta_{0k}$), the general $n$-th order energy correction is defined recursively by projecting the perturbation operator onto the $(n-1)$-th order wave function correction \cite[Sec. 9.2]{levine2014}:
\begin{equation}
E^{(n)} = \langle \Psi_0^{(0)} | \hat{H}^{(1)} | \Psi_0^{(n-1)} \rangle
\end{equation}

Although evaluating the perturbed wave function $\Psi_0^{(n-1)}$ becomes computationally demanding for complex many-body atomic systems, Wigner's $2k+1$ theorem adopted in the perturbation theory in all fundamental quantum mechanical references that is originally formulated in Wigner's book \cite{wigner1959} mathematically optimizes this process by demonstrating that knowledge of the perturbed wave function up to order $k$ is sufficient to calculate the exact energy correction up to order $2k+1$.
\section{Variational Method Using Two Effective Nuclear Charge Parameters Estimation}

\subsection{Theoretical Framework and Trial Wavefunction}
While perturbation theory provides a systematic correction based on the unperturbed nuclear field, it struggles to fully capture the dynamic screening effects introduced by the electronic electric fields \cite[Sec. 1.3]{szabo1982} \cite[Sec. 7]{griffiths2018}. To establish a more accurate upper bound for the ground state energy of the multi-electron structure, the variational method is employed in conjunction with the frozen-core mean field approximation.\cite[Sec. 3]{szabo1982} \cite[Sec. 11]{levine2014}

Because electrons are indistinguishable fermions, the total wavefunction must be completely antisymmetric, which is naturally enforced using a Slater determinant. This formalism inherently accounts for standard Coulomb repulsion and introduces exchange energy (an attractive quantum mechanical correction between electrons with parallel spins). In the  frozen-core mean field approximation method \cite[Sec. 7.1]{bransden1983}, the framework is to freeze the states of two electrons to calculate the average electrostatic potential field (\(V_{\text{eff}}\)) projected onto the third. This approximation approach is assumed for the calculation of potential energies of each electron pair, whose integrals are then solved analytically by exploiting the gamma function. Finally, the results of each of the potential calculations and the kinetic energies are summed up to construct a trial energy function which is then to be minimized in order to find a value for the constant $\alpha$ (i.e. the effective nuclear charge).\\
To implement this minimization mathematically, we construct a trial wave function (Ansatz) featuring a variational parameter ($\alpha$). 
\begin{equation}
E(\alpha) = E_{trial} = \frac{\langle \psi | \hat{H} | \psi \rangle}{\langle \psi | \psi \rangle} \ge E_{ground}
\end{equation}
Physically, ($\alpha$) represents the effective nuclear charge (\(Z_{\text{eff}}\)) felt by the individual electrons due to the shielding effect of the core charge distribution\cite[Sec. 10.8]{levine2014} \cite[Sec. 7]{griffiths2018}. Since \(Z_{\text{eff}}\) is always smaller then the total Z, \(E_{\text{ground}}\) will always be smaller then the \(E_{\text{trial}}\). The used Hamiltonian operator is the same of that in the perturbation part which is given in Equation 3.

\subsection{Trial Wavefunctions and Mathematical Identities}

To proceed with the variational method, we construct hydrogenic trial wavefunctions. To maintain correct SI dimensionality, we explicitly incorporate the Bohr radius ($a_0$) alongside the dimensionless variational parameter $\alpha$ where $\alpha = Z_{\text{eff}}$  \cite{rioux2026}\cite{deng2025}:
\begin{align}
    \psi_{1s}(r) &= \sqrt{\frac{\alpha^3}{\pi a_0^3}} e^{-\alpha r/a_0} \\
    \psi_{2s}(r) &= \sqrt{\frac{\alpha^3}{32\pi a_0^3}} \left( 2 - \frac{\alpha r}{a_0} \right) e^{-\alpha r/(2a_0)}
\end{align}

Because the probability wavefunctions must be normalized over all space, the spherical volume element integrates out the angular dependence:
\begin{equation}
    \int d^3{r} = 4\pi \int_{0}^{\infty} r^2 dr
\end{equation}

For the subsequent radial integrations, we will frequently utilize the standard complete gamma function identity:
\begin{equation}
    \int_{0}^{\infty} r^n e^{-br} dr = \frac{n!}{b^{n+1}}
\end{equation}

\subsection{Kinetic Energy Expectation Values \(\langle T \rangle\)}

The kinetic energy expectation value in SI units is evaluated using the operator \(\hat{T} = -\frac{\hbar^2}{2m_e}\nabla^2\), which can be alternatively expressed in its symmetric gradient form:
\begin{equation}
    \langle T \rangle = \frac{\hbar^2}{2m_e} \int |\nabla \psi|^2 d^3{r}
\end{equation}

Let us evaluate this for the \(1s\) orbital. The radial gradient is:
\begin{equation}
    \nabla \psi_{1s} = \frac{\partial \psi_{1s}}{\partial r} = -\frac{\alpha}{a_0} \sqrt{\frac{\alpha^3}{\pi a_0^3}} e^{-\alpha r/a_0}
\end{equation}

Substituting this into the kinetic energy integral yields:
\begin{equation}
\begin{aligned}
    \langle T_{1s} \rangle &= \frac{\hbar^2}{2m_e} (4\pi) \int_{0}^{\infty} \left( -\frac{\alpha}{a_0} \sqrt{\frac{\alpha ^3}{\pi a_0^3}} \right)^2 e^{-2\alpha r/a_0} r^2 dr \\
    &= \frac{2\hbar^2 \alpha^5}{m_e a_0^5} \int_{0}^{\infty} r^2 e^{-2\alpha r/a_0} dr
\end{aligned}
\end{equation}

Applying the gamma integral (equation 72) identity with \(n=2\) and \(b=2\alpha/a_0\):
\begin{equation}
    \langle T_{1s} \rangle = \frac{2\hbar^2 \alpha^5}{m_e a_0^5} \left( \frac{2!}{(2\alpha/a_0)^3} \right) = \frac{\hbar^2}{2m_e a_0^2} \alpha^2
\end{equation}

Recognizing that the physical constant group \(\frac{\hbar^2}{2m_e a_0^2}\) represents exactly one Rydberg of energy (\(\approx 13.6057\ \text{eV}\)), we can express the kinetic energy expectations directly in electron-volts. Using the general principal quantum number dependence (\(\langle T \rangle \propto 1/n^2\)), the kinetic energies for the \(n=1\) and \(n=2\) states are:
\begin{equation}
    \langle T_{1s} \rangle = \frac{a^2}{1^2}(13.6057\ \text{eV}) = \alpha^2 (13.6057\ \text{eV})
\end{equation}
\begin{equation}
    \langle T_{2s} \rangle = \frac{\alpha^2}{2^2} (13.6057\ \text{eV}) = \frac{\alpha^2}{4} (13.6057\ \text{eV})
\end{equation}
which align with the result in \cite[Sec. 2.32]{rioux2026}.
In the Hamiltonian of multi-electron atoms, there is only one kinetic energy expectation value for each orbital because the nucleus is static. However, there are two potential terms that must be physically separated because the electron-nucleus potential energy \(V_{N}\) represents the attractive interaction binding electrons to the nucleus, whereas the electron-electron repulsion \(V_{ee}\) accounts for the opposing repulsive force between the electrons themselves. Including both terms is absolutely essential to accurately calculate the total energy of the system and to properly model the shielding effect that the electrons in the lithium atom exert on one another.

\subsection{Nucleus Potential Energy \(\langle V_{N} \rangle\)}

The Coulomb interaction between the electrons and the nucleus (\(Z=3\)) in SI units includes the potential operator \(-\frac{Z e^2}{4\pi\varepsilon_0 r}\). Evaluating the expectation value for the \(1s\) orbital:
\begin{equation}
\begin{aligned}
    \langle V_{N, 1s} \rangle &= 4\pi \int_{0}^{\infty} \left( \sqrt{\frac{\alpha^3}{\pi a_0^3}} e^{-\alpha r/a_0} \right) \left(-\frac{3e^2}{4\pi\varepsilon_0 r}\right) \left( \sqrt{\frac{\alpha^3}{\pi a_0^3}} e^{-\alpha r/a_0} \right) r^2 dr \\
    &= -\frac{12 e^2 \alpha^3}{4\pi\varepsilon_0 a_0^3} \int_{0}^{\infty} r e^{-2\alpha r/a_0} dr \\
    &= -\frac{3 e^2 \alpha^3}{\pi\varepsilon_0 a_0^3} \left( \frac{1!}{(2\alpha/a_0)^2} \right) = -\frac{3 e^2 \alpha^3}{\pi\varepsilon_0 a_0^3} \left( \frac{a_0^2}{4\alpha^2} \right) = -3\alpha \left( \frac{e^2}{4\pi\varepsilon_0 a_0} \right)
\end{aligned}
\end{equation}

The physical constant group \(\left( \frac{e^2}{4\pi\varepsilon_0 a_0} \right)\) represents exactly 2 Rydbergs (\(\approx 27.2114\ \text{eV}\)). Thus, the potential energy can be expressed directly in eV:
\begin{equation}
    \langle V_{N, 1s} \rangle = -3 \alpha(27.2114\ \text{eV}) = 2(-3 \alpha)(1\ \text{Rydberg)} = -6 \alpha (13.6057\text{eV})
\end{equation}

Similarly, for the \(2s\) orbital, using the general scaling relation \(\langle V \rangle \propto -\frac{Z \cdot Z_{\text{eff}}}{n^2}\):
\begin{equation}
    \langle V_{N, 2s} \rangle = -\frac{3 \alpha}{4} \left( \frac{e^2}{4\pi\varepsilon_0 a_0} \right) = 2(-\frac {3}{4} \alpha) (1\ \text{Rydberg}) =- \frac{3}{2} \alpha(13.6057\ \text{eV})
\end{equation}

\subsection{Electron-Electron Repulsion \(\langle V_{ee} \rangle\)}

For the inter-electronic repulsion, we utilize the multipole expansion where \(\frac{1}{|{r}_1 - {r}_2|} \rightarrow \frac{1}{r_>}\), with \(r_> \equiv \max(r_1, r_2)\). The SI operator is \(\frac{e^2}{4\pi\varepsilon_0 r_{12}}\).

\begin{equation}
\begin{aligned}
    \langle V_{1s, 1s} \rangle &= \frac{e^2}{4\pi\varepsilon_0} (4\pi)^2 \left( \frac{ \alpha^3}{\pi a_0^3} \right)^2 \int_{0}^{\infty} r_1^2 e^{-2 \alpha r_1/a_0} \Bigg[ \int_{0}^{r_1} \frac{1}{r_1} r_2^2 e^{-2 \alpha r_2/a_0} dr_2 \\
    &\quad + \int_{r_1}^{\infty} \frac{1}{r_2} r_2^2 e^{-2 \alpha r_2/a_0} dr_2 \Bigg] dr_1
\end{aligned}
\end{equation}

Applying integration by parts for the terms inside the brackets (which physically represents calculating the classical electrostatic potential of a spherical charge cloud), we obtain:
\begin{equation}
    \text{Inner Bracket} = \frac{a_0^3}{4 \alpha ^3 r_1} \left[ 1 - e^{-2 \alpha r_1/a_0} \left(1 + \frac{\alpha r_1}{a_0}\right) \right]
\end{equation}

Substituting this back into the main outer integral yields:
\begin{equation}
\begin{aligned}
    \langle V_{1s, 1s} \rangle &= \frac{e^2}{4\pi\varepsilon_0} \left(\frac{4 \alpha ^3}{a_0^3}\right) \int_{0}^{\infty} \left( r_1 e^{-2 \alpha r_1/a_0} - r_1 e^{-4 \alpha r_1/a_0} - \frac{\alpha}{a_0} r_1^2 e^{-4 \alpha r_1/a_0} \right) dr_1 \\
    &= \frac{e^2}{4\pi\varepsilon_0} \left(\frac{4 \alpha ^3}{a_0^3}\right) \left[ \frac{1!}{(2 \alpha /a_0)^2} - \frac{1!}{(4 \alpha /a_0)^2} - \frac{\alpha}{a_0}\frac{2!}{(4 \alpha /a_0)^3} \right] \\
    &= \frac{e^2}{4\pi\varepsilon_0} \left(\frac{4 \alpha ^3}{a_0^3}\right) \left[ \frac{a_0^2}{4 \alpha ^2} - \frac{a_0^2}{16 \alpha ^2} - \frac{2a_0^2}{64 \alpha ^2} \right] \\
    &= \frac{e^2}{4\pi\varepsilon_0 a_0} (4\alpha ) \left( \frac{5}{32} \right) = \frac{5}{8}\alpha \left( \frac{e^2}{4\pi\varepsilon_0 a_0} \right) \\
    &= 2(\frac{5}{8}\alpha ) (1\ \text{Rydberg}) = \frac{5}{4}\alpha  (13.6057\ \text{eV})
\end{aligned}
\end{equation}

By equivalent spatial integration methods, the Coulomb integral for the \(1s\) and \(2s\) interaction evaluates to:
\begin{equation}
    \langle V_{1s, 2s} \rangle = \frac{17}{81}\alpha  \left( \frac{e^2}{4\pi\varepsilon_0 a_0} \right) = 2(\frac{17}{81}\alpha) (1\ \text{Rydberg}) = \frac{34}{81}\alpha (13.6057\ \text{eV})
\end{equation}

\subsection{Total Energy and Minimization}

The total ground-state energy of the \(1s^2 2s^1\) configuration is the linear combination of its nine constituent expectation values: two kinetic terms and two nuclear potential terms for the \(1s\) core, one kinetic and one nuclear potential term for the \(2s\) valence electron, one core-core (\(1s-1s\)) repulsion term, and two core-valence (\(1s-2s\)) repulsion terms.

Factoring out the standard Rydberg energy scale, the parametric energy function \(E(\alpha)\) in SI units is (factoring out the Rydberg constant from all terms) assembled as:
\begin{equation}
\begin{aligned}
    E(\alpha) &= 2\langle T_{1s} \rangle + \langle T_{2s} \rangle + 2\langle V_{N, 1s} \rangle + \langle V_{N, 2s} \rangle + \langle V_{1s, 1s} \rangle + 2\langle V_{1s, 2s} \rangle \\
    E(\alpha) &= 2\left(\alpha^2\right) + \frac{\alpha^2}{4} + 2(-6\alpha) + \left(-\frac{3\alpha}{2}\right) + \frac{5\alpha}{4} + 2\left(\frac{34\alpha}{81}\right) \\
    E(\alpha) &= \frac{9\alpha^2}{4} - \frac{3697}{324}\alpha
\end{aligned}
\end{equation}

Since there are 2 electrons in 1s orbital, the energies \(T_{1s}\) and \(V_{N,1s}\) are multiplied by a factor of 2, and the same is correct for \(V_{1s,2s}\) because each electron in the 1s orbital interact with the electron in 2s orbital. 

According to the variational principle, this function represents an upper bound to the true ground-state energy. To find the optimal effective nuclear charge that minimizes the system's energy, we differentiate \(E(\alpha)\) with respect to the variational parameter \(\alpha\) and set it to zero:
\begin{equation}
    \frac{dE}{d\alpha} = \frac{9}{2}\alpha - \frac{3697}{324} = 0 \implies \alpha \approx 2.536
\end{equation}

This value indicates that the \(1s\) electrons experience an effective nuclear charge of approximately \(+2.536e\), explicitly demonstrating the shielding effect. Substituting the optimized \(\alpha\) parameter back into the energy equation yields the variational minimum energy limit, which we convert directly to electron-volts:
\begin{equation}
\begin{aligned}  
    E_{\text{min}} = \left[ \frac{9}{4}(2.536)^2 - \frac{3697}{324}(2.536) \right](1\ \text{Rydberg})\\
    \approx -14.467 (13.6057\ \text{eV}) \approx {-196.83\ \text{eV}}
\end{aligned}
\end{equation}

The non-relativistic reference limit for the lithium ground state is approximately \(-203.5\ \text{eV}\) \cite{nist2024}. The simple single-parameter variational ansatz captures the energy to within \(\sim 3.3\%\) accuracy, confirming its effectiveness despite the absence of dynamic angular correlation.

It should be emphasized that the single-parameter variational framework presented in this study operate strictly within the non-relativistic regime. While this approach effectively models the primary electronic screening and correlation effects, achieving true spectroscopic precision necessitates the inclusion of relativistic and quantum electrodynamic (QED) corrections which were not considered in this study. For a recent and comprehensive theoretical treatment incorporating these relativistic effects into the ground-state calculation of lithium using single-parameter variational approach, the reader is referred to the detailed framework presented in \cite{deng2025} which obtains an error percentage of \(\sim 2.5\%\) that is closer to the reference value than the result obtained here \(\sim 3.3\%\). 

\subsection{Two Parameter Estimation}
Although one parameter estimation yielded a result 3.3\% close to the reference result, it is more proper to expand the estimation to a second parameter. That is even more reasonable because of the two orbital nature of the Lithium atom. For this, let the previously calculated shielding parameter be $\alpha$, that is the shielding effect felt by electrons in the 1s orbital. Therefore we introduce $\beta$ which is the shielding effect felt in the 2s orbital. \cite[Sec. 2.32]{rioux2026} \cite[Sec. 31]{wilson1935}

The calculation algorithm of a second shielding parameter follows the same steps as that of the first one. The difference is seen, however, in the definition of the hydrogenic wave functions of the orbitals. Specifically in the definition of the 2s orbital's wave function where $\beta$ is used instead of $\alpha$, and is defined as $\beta = Z-1$ where the total atomic number $Z$ is lowered significantly by a factor of 1 to add even more shielding effect due to the 1s electrons. The wave functions are defined as:

\begin{equation}
    \psi(1s) = \sqrt{\frac{\alpha^3}{\pi}} e^{-\alpha r}
\end{equation}

where $\psi(1s)$ represents the spatial distribution of the core electrons, with $\alpha$ acting as the optimized effective charge under core-core screening.
\begin{equation}
    \psi(2s) = \sqrt{\frac{\beta^3}{32\pi}} (2 - \beta r) e^{-\frac{\beta r}{2}}
\end{equation}
where $\psi(2s)$ denotes the wave function of the valence electron, modulated by $\beta$ to simulate the robust shielding provided by the underlying $1s^2$ shell.

By evaluating the corresponding Hamiltonian matrix elements (factoring out the Rydberg energy unit from all the terms) using these non-orthogonal trial states within the Slater determinant framework, the individual kinetic, potential, Coulomb, and exchange expectation values are explicitly derived as:

\begin{description}
    \item[Core Kinetic Energy ($T_{1s}$):] Represents the expectation value of the kinetic energy for a single electron residing in the inner $1s$ shell:
    \begin{equation}
        \langle T_{1s} \rangle  = \frac{\alpha^2}{2}
    \end{equation}

    \item[Valence Kinetic Energy ($T_{2s}$):] Describes the expectation value of the kinetic energy for the outer-shell $2s$ electron:
    \begin{equation}
        \langle T_{2s} \rangle  = \frac{\beta^2}{8}
    \end{equation}

    \item[Core Electron-Nucleus Potential ($V_{N,1s}$):] Quantifies the attractive Coulomb potential energy between the nucleus of charge $Z$ and a core $1s$ electron:
    \begin{equation}
        \langle V_{N,1s} \rangle = -Z\alpha
    \end{equation}

    \item[Valence Electron-Nucleus Potential ($V_{N,2s}$):] Represents the attractive Coulomb potential energy acting on the valence $2s$ electron:
    \begin{equation}
        \langle V_{N,2s}(\beta) \rangle = \frac{-Z\beta}{4}
    \end{equation}

    \item[Core-Core Coulomb Repulsion ($V_{1s1s}$):] Accounts for the direct electrostatic repulsion energy between the two spin-paired electrons sharing the spatial $1s$ orbital:
    \begin{equation}
        \langle V_{1s1s}(\alpha) \rangle = \frac{5}{8}\alpha
    \end{equation}

    \item[Direct Core-Valence Repulsion ($V_{1s2s}$):] Represents the classical direct Coulomb repulsion integral between a core $1s$ electron and the valence $2s$ electron:
    \begin{equation}
       \langle V_{1s2s}(\alpha, \beta) \rangle = \alpha\beta \frac{\beta^4 + 10\alpha\beta^3 + 8\alpha^4 + 20\alpha^3\beta + 12\alpha^2\beta^2}{(2\alpha + \beta)^5}
    \end{equation}

    \item[Kinetic Energy Exchange Contribution ($T_{1s2s}$):] A non-classical kinetic term that arises because the $1s$ and $2s$ spatial wave functions are no longer strictly orthogonal when governed by different screening parameters ($\alpha \neq \beta$):
    \begin{equation}
        \langle T_{1s2s}(\alpha, \beta) \rangle = -4\sqrt{2}\alpha^{\frac{5}{2}}\beta^{\frac{5}{2}} \frac{\beta - 4\alpha}{(2\alpha + \beta)^4}
    \end{equation}
    This particular term is worth noting as it is non-existing in the case of one shielding parameter. That is because when the shielding effect is reduced to one parameter only inter-orbital interaction converges to zero due to the orthogonality of the wave functions of both orbitals.   
    \item[Core-Valence Exchange Repulsion ($V_{12,12}$):] This inter-electronic exchange integral represents the energy correction due to the Pauli exclusion principle as the electrons have parallel spin states(both electrons have a spin state of 2 as shown in the subscript):
    \begin{equation}
        \langle V_{12,12}(\alpha, \beta) \rangle = 16\alpha^3\beta^3 \frac{13\beta^2 + 20\alpha^2 - 30\beta\alpha}{(\beta + 2\alpha)^7}
    \end{equation}

    \item[Spatial Overlap Integral ($S_{1s,2s}$):] is a measurment of  the actual spatial non-orthogonality \cite[Sec. 7.3]{griffiths2018} which is due to the wave function mixing between both orbitals:
    \begin{equation}
        \langle S_{1s2s}(\alpha, \beta) \rangle = 32\sqrt{2}\alpha^{\frac{3}{2}}\beta^{\frac{3}{2}} \frac{\alpha - \beta}{(2\alpha + \beta)^4}
    \end{equation}  
\end{description}
The next step is to sum the separate energies to find the trial energy as a function of both $\alpha$ and $\beta$:

\begin{equation}
\langle E(\alpha, \beta) \rangle = \frac{
\begin{array}{l}
2\langle T_{1s}(\alpha) \rangle + \langle T_{2s}(\beta) \rangle - \langle T_{1s}(\alpha) \rangle S_{1s2s}(\alpha, \beta)^2 - 2\langle T_{1s2s}(\alpha, \beta) \rangle S_{1s2s}(\alpha, \beta)... \\
+2\langle V_{N1s}(\alpha) \rangle + \langle V_{N2s}(\beta) \rangle - \langle V_{N1s}(\alpha) \rangle S_{1s2s}(\alpha, \beta)^2 - 2\langle V_{N1s2s}(\alpha, \beta) \rangle S_{1s2s}(\alpha, \beta)... \\
+2\langle V_{1s2s}(\alpha, \beta) \rangle + \langle V_{1s1s}(\alpha) \rangle - 2\langle V_{1112}(\alpha, \beta) \rangle S_{1s2s}(\alpha, \beta) - \langle V_{1212}(\alpha, \beta) \rangle
\end{array}
}{1 - S_{1s2s}(\alpha, \beta)^2}
\end{equation}

The same reasoning used in the previous parameter estimation methodology for multiplying some of the terms by a factor of 2 is also applied here, particularly to account for the presence of two electrons in the 1s orbital, and thus 2 pairs with the electron in the 2s orbital each. The term in the denominator is simply to ensure the normalization of the results of the total energy expectation value.

The final step is minimizing the trial energy expectation value, and that follows the same logic as before the substituting with the values of $\alpha$ and $\beta$ in the formula of $\langle E(\alpha, \beta) \rangle$, yielding (putting back the 2 Rydberg energy units):

\begin{gather}
\text{Given} \quad \frac{\partial}{\partial\alpha}\langle E(\alpha, \beta) \rangle = 0 \quad \frac{\partial}{\partial\beta}\langle E(\alpha, \beta) \rangle = 0 \\[10pt]
\begin{pmatrix} \alpha \\ \beta \end{pmatrix} = \text{Find}(\alpha, \beta) \quad 
\begin{pmatrix} \alpha \\ \beta \end{pmatrix} = \begin{pmatrix} 2.6797 \\ 1.8683 \end{pmatrix} \quad \\[10pt]
\langle E(\alpha, \beta) \rangle = -201.187 \text{ eV} 
\end{gather}
The reference value is $E_{ref} = -203.5\ \text{eV}$ \cite{nist2024}. The error percentage then can be calculated as:
\begin{equation}
 \left| \frac{\langle E(\alpha, \beta) \rangle - E_{ref}}{E_{ref}} \right| = 1.13\%
\end{equation}
The results all agree with those in \cite{rioux2026}.

\section{Conclusion}
In this study, the ground-state energy of the lithium atom ($1s^2 2s^1$) was systematically investigated by deploying and contrasting two foundational pillars of quantum mechanics: Rayleigh-Schrödinger perturbation theory and the variational method. By rigorously enforcing the Pauli exclusion principle via a totally antisymmetric Slater determinant framework, the non-interacting unperturbed baseline energy was established at $-275.51\,\text{eV}$. 

First-order perturbation theory successfully incorporated the averaged electrostatic inter-electronic repulsion through the analytical evaluation of static Coulomb and quantum exchange integrals, providing a massive first-order upward shift to $-192.01\,\text{eV}$. To account for dynamic electron correlation, higher-order virtual single-electron excitation channels were computationally evaluated. Resolving the spherically symmetric virtual transitions from the $3s$ up to the $6s$ Rydberg shells yielded a cumulative second-order energy correction of $-4.35\,\text{eV}$, culminating in a total perturbative ground-state energy of $-196.36\,\text{eV}$.\\
Fig.~\ref{fig:2} illustrates the radial probability density functions for the $1s$ and $2s$ (as given in Appendix A.1) orbitals across sequential perturbation treatments along with a curve produced using the fitted parameter obtained upon applying the variational approach for each orbital to be used as a reference point as it is the closest obtained result to the reference energy value($-203.5$ eV). Comparing the perturbation order correction curves of the same orbital, it is clearly seen that the zeroth and first-order profiles exhibit highly localized, unshielded hydrogen-like behaviors, while the second-order corrected curves visibly shift closer to realistic multi-body configurations where the probability is more distributed over space. Comparing the curves of both orbitals from different perturbation orders, it is notable that the valence $2s$ orbital distribution becomes broader and exhibits a noticeable outward spatial relaxation compared to the $1s$ curves, demonstrating that the perturbation formalism successfully captures effective physical shielding profiles and electron-electron avoidance configurations without the explicit apriori inclusion of variational shielding parameters into the calculation.

\begin{figure}[h]
    \centering
    \includegraphics[scale=0.35]{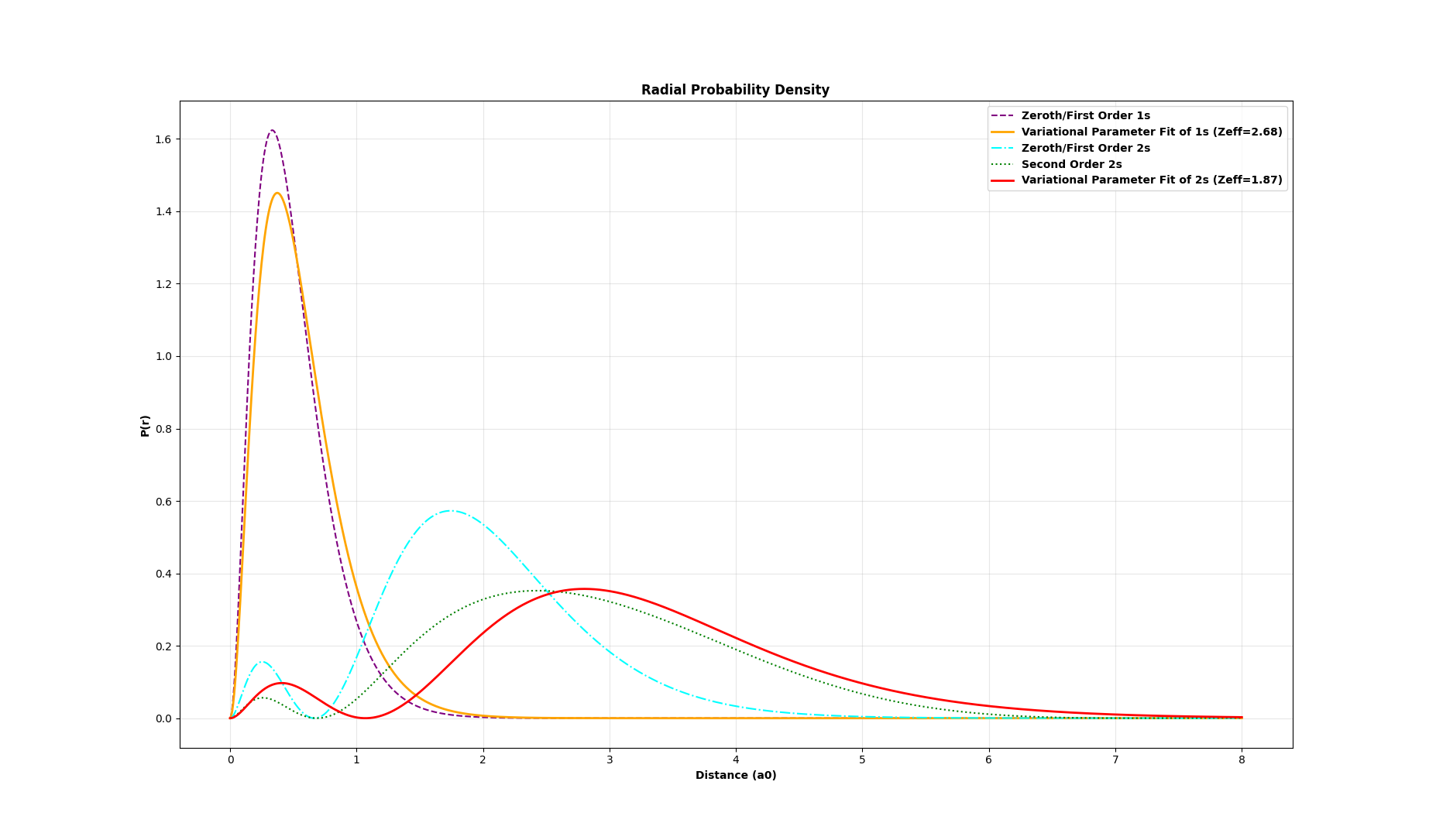}
\caption{Radial probability density functions $P(r)$ for the 1s and 2s shells of the lithium atom, comparing all evaluated frameworks in this study: 0th/1st and 2nd-order perturbation theories against the single- and two-parameter variational methods utilizing optimized effective charges \cite{github}.}
    \label{fig:2}
\end{figure}
To better capture the non-linear relaxation and radial screening effects introduced by the electronic charge clouds, a non-orthogonal multi-parameter variational approach was subsequently implemented. By introducing independent variational shielding parameters for the inner and outer shells, this model explicitly map the differential screening landscapes of the atom. Minimizing the global energy functional isolated the optimal effective nuclear charges at $\alpha = 2.6797$ for the $1s$ core and $\beta = 1.8683$ for the $2s$ valence electron. This multi-parameter optimization successfully established a highly accurate upper energy bound of $-201.187\,\text{eV}$.

\begin{figure}[h]
    \centering
    \includegraphics[width=0.86\linewidth]{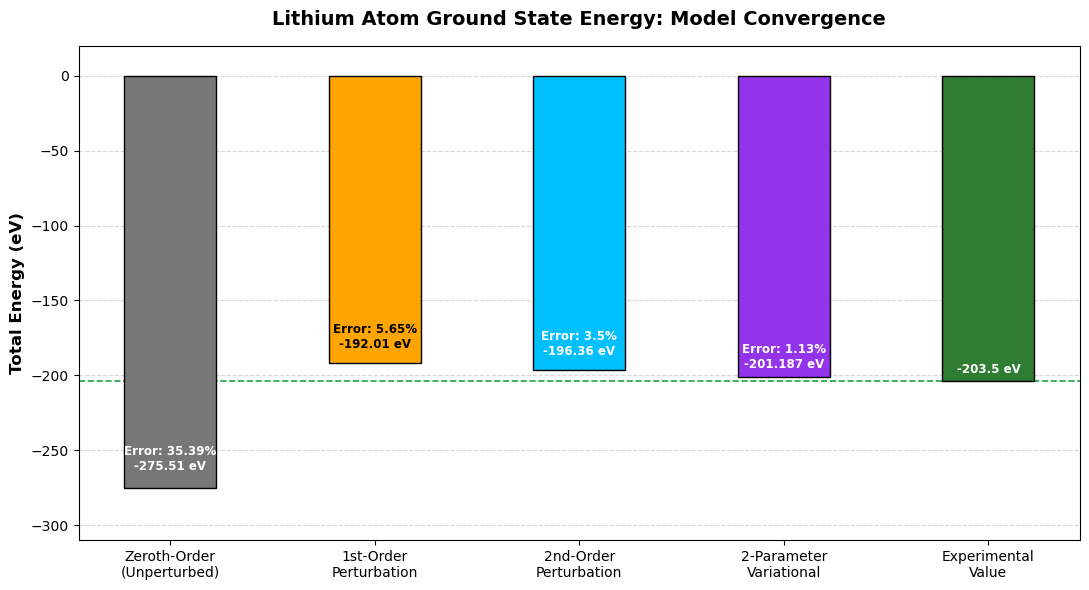}
    \caption{Comprehensive comparison of total ground-state energy convergence across sequential perturbation orders and the multi-parameter variational method against the non-relativistic reference baseline ($-203.5\,\text{eV}$) \cite{nist2024}. Percentage values indicate the relative error calculated with respect to the reference baseline.}
    \label{fig:3}
\end{figure}

A definitive comparative insight emerges when evaluating these advanced theoretical frameworks against the non-relativistic reference ground-state benchmark of approximately $-203.5\,\text{eV}$\cite{nist2024}, as visually summarized in Fig.~\ref{fig:3}. The multi-parameter variational framework outpaced the second-order single-excitation perturbative treatment, shrinking the relative error margin to a mere $1.13\%$, whereas the second-order perturbation expansion converged to a $3.51\%$ relative error. 

The superior accuracy of the variational approach is directly attributed to its capacity to dynamically adjust the underlying orbital dimensions to simulate physical shielding. Rather than experiencing the bare electrostatic force of the unshielded nucleus ($Z=3$), the inner core and outer valence electrons adjust to effective atomic numbers of $Z_{\text{eff},1s} = 2.6797$ and $Z_{\text{eff},2s} = 1.8683$, respectively. 

Finally, the residual energy gap between the optimized variational limit and the reference baseline highlights the persistent frontiers of many-body atomic physics. This remaining discrepancy is predominantly driven by angular electron correlation effects—such as simultaneous multi-particle double-excitations into non-spherical ($p^2, d^2$) spatial manifolds—alongside minor relativistic and fine-structure corrections. Ultimately, both frameworks successfully demonstrate how systematic mathematical refinements can transform a crude independent-particle approximation into a highly predictive model of multi-electron atomic structures.

\appendix

\section{Detailed Evaluation of Coulomb and Exchange Integrals}

\subsection{Radial Functions and Corresponding Probability Distributions}
The standard unperturbed hydrogenic single-particle radial functions for a general bare nuclear charge $Z$ are defined below, where the characteristic atomic length scale is governed by the Bohr radius ($a_0$) and energies scale in terms of the Rydberg constant ($1\,\text{Ry} \approx 13.6057\ \text{eV}$) \cite[Sec. 6.6]{levine2014}:
\begin{align}
R_{1s}(r) &= 2\left(\frac{Z}{a_0}\right)^{3/2}e^{-\frac{Zr}{a_0}} \\
R_{2s}(r) &= \frac{1}{2\sqrt{2}}\left(\frac{Z}{a_0}\right)^{3/2}\left(2-\frac{Zr}{a_0}\right)e^{-\frac{Zr}{2a_0}}
\end{align}
The corresponding radial probability distributions, mapping the isotropic spatial profiles of the shells, are given by \cite[Sec. 6.6]{levine2014}:
\begin{align}
P_{1s}(r) &= r^2|R_{1s}(r)|^2 \\
P_{2s}(r) &= r^2|R_{2s}(r)|^2
\end{align}
 The python code and its detailed description provided in \cite{github} demonstrate the algorithm used to compute the radial probability density functions and plot the results as a function of the distance from the nucleus for each perturbation correction as well as the variational approximation in order to provide further insight regarding the frameworks used in this study.

\subsection{General Radial Reduction via Multipole Expansion}
The inter-electronic Coulomb repulsion operator $1/r_{12} = 1/|\mathbf{r}_1 - \mathbf{r}_2|$ can be fundamentally represented using the standard spherical multipole expansion:
\begin{equation}
\frac{1}{|\mathbf{r}_1 - \mathbf{r}_2|} = \sum_{l=0}^{\infty} \sum_{m=-l}^{l} \frac{4\pi}{2l+1} \frac{r_{<}^{l}}{r_{>}^{l+1}} Y_{lm}^*(\Omega_1) Y_{lm}(\Omega_2)
\end{equation}
For multi-electron systems confined strictly to spherically symmetric $s$-orbitals ($l=0, m=0$), the angular integrations over the solid angles ($d\Omega_1 d\Omega_2$) isolate the monopole term exclusively. Due to the geometric orthogonality of spherical harmonics, all higher-order angular channels ($l \ge 1$) vanish identically. Thus, the local repulsion operator collapses strictly to its isotropic monopole bound, $1/r_{>}$, where $r_{>} \equiv \max(r_1, r_2)$ and $r_{<} \equiv \min(r_1, r_2)$ \cite[Sec. 6.5]{levine2014}.

Splitting the radial domain to accommodate this piecewise boundary conditions transforms the generic spatial double integral for inter-electronic interactions into a structured two-region functional layout:
\begin{equation}
J_{ab} = \int_0^\infty r_1^2 \rho_a(r_1) \left[ \frac{1}{r_1} \int_0^{r_1} r_2^2 \rho_b(r_2)\,dr_2 + \int_{r_1}^\infty r_2 \rho_b(r_2)\,dr_2 \right] dr_1
\end{equation}
Physically, this represents the classical electrostatic potential energy separating two localized charge clouds. In direct harmony with Gauss's Law, the bracketed term dictates the effective potential field generated by electron $b$: the charge density enclosed within the interior sphere of radius $r_1$ acts as a centralized point charge, whereas the exterior distribution acts as a uniform potential shell.

\subsection{Explicit Derivation of the $1s\text{–}1s$ Coulomb Integral ($J_{1s,1s}$)}
To evaluate the static repulsion within the core shell, we isolate the $1s\text{–}1s$ Coulomb expectation value. Let the local effective potential generated by the $1s$ density cloud be designated as $U_{1s}(r_1)$:
\begin{equation}
U_{1s}(r_1) = \mathcal{I}_1(r_1) + \mathcal{I}_2(r_1) = \frac{1}{r_1} \int_{0}^{r_1} r_2^2 \rho_{1s}(r_2) \, dr_2 + \int_{r_1}^{\infty} r_2 \rho_{1s}(r_2) \, dr_2
\end{equation}
To streamline the evaluation, we define the scaled scaling variable $\alpha = 2Z/a_0$, simplifying the core density expression to $\rho_{1s}(r) = \frac{\alpha^3}{2} e^{-\alpha r}$.

\textbf{Step 1: Evaluation of the Interior Integral $\mathcal{I}_1(r_1)$}\\
Applying standard integration by parts to the interior radial boundary yields:
\begin{equation}
\mathcal{I}_1(r_1) = \frac{1}{r_1} \int_{0}^{r_1} r_2^2 \left( \frac{\alpha^3}{2} e^{-\alpha r_2} \right) dr_2 = \frac{1}{r_1} \left[ 1 - e^{-\alpha r_1} \left( \frac{\alpha^2 r_1^2}{2} + \alpha r_1 + 1 \right) \right]
\end{equation}

\textbf{Step 2: Evaluation of the Exterior Integral $\mathcal{I}_2(r_1)$}\\
Similarly, evaluating the exterior shell component from $r_1$ to infinity gives:
\begin{equation}
\mathcal{I}_2(r_1) = \int_{r_1}^{\infty} r_2 \left( \frac{\alpha^3}{2} e^{-\alpha r_2} \right) dr_2 = e^{-\alpha r_1} \left( \frac{\alpha^2 r_1}{2} + \frac{\alpha}{2} \right)
\end{equation}

\textbf{Step 3: Construction of the Global Core Potential $U_{1s}(r_1)$}\\
Summing the two sub-domains triggers an algebraic cancellation of the linear spatial terms, leaving a compact, closed-form classical screening field:
\begin{equation}
U_{1s}(r_1) = \mathcal{I}_1(r_1) + \mathcal{I}_2(r_1) = \frac{1}{r_1} - e^{-\alpha r_1} \left( \frac{\alpha}{2} + \frac{1}{r_1} \right)
\end{equation}

\textbf{Step 4: Final Integration of $J_{1s,1s}$}\\
Projecting this effective electrostatic potential back across the outer core distribution yields:
\begin{equation}
\begin{aligned}
J_{1s,1s} &= \int_{0}^{\infty} r_1^2 \left( \frac{\alpha^3}{2} e^{-\alpha r_1} \right) \left[ \frac{1}{r_1} - e^{-\alpha r_1} \left( \frac{\alpha}{2} + \frac{1}{r_1} \right) \right] dr_1 \\
&= \frac{\alpha^3}{2} \left[ \int_{0}^{\infty} r_1 e^{-\alpha r_1} \,dr_1 - \int_{0}^{\infty} e^{-2\alpha r_1} \left( \frac{\alpha r_1^2}{2} + r_1 \right) \,dr_1 \right]
\end{aligned}
\end{equation}
Evaluating these standard continuous definitive channels via the Euler gamma identity (Equation 72) resolves to:
\begin{equation}
J_{1s,1s} = \frac{\alpha^3}{2} \left[ \frac{1}{\alpha^2} - \frac{1}{8\alpha^2} - \frac{1}{4\alpha^2} \right] = \frac{\alpha^3}{2} \left( \frac{5}{8\alpha^2} \right) = \frac{5\alpha}{16}
\end{equation}
Restoring the physical variables ($\alpha = 2Z/a_0$) maps the final $1s\text{–}1s$ Coulomb repulsion energy strictly as a function of the core nuclear boundary:
\begin{equation}
J_{1s,1s} = \frac{5Ze^2}{32\pi \epsilon_0 a_0} \quad \xrightarrow{\text{for } Z=3} \quad J_{1s,1s} \approx 51.02\ \text{eV}
\end{equation}

\subsection{Explicit Derivation of the $1s\text{–}2s$ Coulomb Integral ($J_{1s,2s}$)}
To solve the inter-shell interaction $J_{1s,2s}$, the effective monopole potential fields are mapped over the nodally complex $2s$ orbital density distribution:
\begin{equation}
U_{2s}(r_1) = \frac{1}{r_1} \int_0^{r_1} r_2^2 \rho_{2s}(r_2)\,dr_2 + \int_{r_1}^\infty r_2 \rho_{2s}(r_2)\,dr_2
\end{equation}
Performing systematic integration by parts over the multi-termed polynomial array of the $2s$ shell structures yields the complete spatial screening field:
\begin{equation}
U_{2s}(r_1) = \frac{1}{r_1} - e^{-\frac{Zr_1}{a_0}} \left( \frac{1}{r_1} + \frac{3Z}{4a_0} + \frac{Z^2r_1}{4a_0^2} + \frac{Z^3 r_1^2}{8a_0^3} \right)
\end{equation}
Next, this comprehensive potential profile is evaluated across the inner core density $\rho_{1s}(r_1)$:
\begin{equation}
J_{1s,2s} = 4\left( \frac{e^2}{4 \pi \epsilon_0}\right)\left(\frac{Z}{a_0}\right)^3 \int_0^\infty \left[ r_1 e^{-\frac{2Zr_1}{a_0}} - e^{-\frac{3Zr_1}{a_0}} \left( r_1 + \frac{3Z}{4a_0}r_1^2 + \frac{Z^2}{4a_0^2}r_1^3 + \frac{Z^3}{8a_0^3}r_1^4 \right) \right] \,dr_1
\end{equation}
Invoking the gamma identity independently for each unique polynomial channel yields:
\begin{align}
\int_0^\infty r_1 e^{-\frac{2Zr_1}{a_0}} \,dr_1 &= \frac{a_0^2}{4Z^2} \\
\int_0^\infty r_1 e^{-\frac{3Zr_1}{a_0}} \,dr_1 &= \frac{a_0^2}{9Z^2} \\
\frac{3Z}{4a_0} \int_0^\infty r_1^2 e^{-\frac{3Zr_1}{a_0}} \,dr_1 &= \frac{3Z}{4a_0} \left( \frac{2a_0^3}{27Z^3} \right) = \frac{a_0^2}{18Z^2} \\
\frac{Z^2}{4a_0^2} \int_0^\infty r_1^3 e^{-\frac{3Zr_1}{a_0}} \,dr_1 &= \frac{Z^2}{4a_0^2} \left( \frac{6a_0^4}{81Z^4} \right) = \frac{a_0^2}{54Z^2} \\
\frac{Z^3}{8a_0^3} \int_0^\infty r_1^4 e^{-\frac{3Zr_1}{a_0}} \,dr_1 &= \frac{Z^3}{8a_0^3} \left( \frac{24a_0^5}{243Z^5} \right) = \frac{a_0^2}{81Z^2}
\end{align}
Summing the collective exterior screened contributions results in:
\begin{equation}
\frac{a_0^2}{Z^2} \left( \frac{1}{9} + \frac{1}{18} + \frac{1}{54} + \frac{1}{81} \right) = \frac{16a_0^2}{81Z^2}
\end{equation}
Subtracting this grouped valuation from the localized lead channel isolates the analytical solution:
\begin{equation}
J_{1s,2s} = 4\left(\frac{e^2}{4 \pi \epsilon_0}\right)\left(\frac{Z}{a_0}\right)^3 \left[ \frac{a_0^2}{4Z^2} - \frac{16a_0^2}{81Z^2} \right] = \frac{17Z}{81a_0}\left(\frac{e^2}{4 \pi \epsilon_0}\right)
\end{equation}
Evaluating this final expression for the lithium baseline ($Z=3$) yields:
\begin{equation}
J_{1s,2s} = \frac{17(3)}{81a_0}\left(\frac{e^2}{4 \pi \epsilon_0}\right) \approx 17.135\ \text{eV} \implies 2J_{1s,2s} \approx 34.27\ \text{eV}
\end{equation}
This closed-form mathematical expression aligns perfectly with standard quantum chemistry reference literature \cite[Sec. 10.5]{levine2014}.

\subsection{Explicit Derivation of the Exchange Integral ($K_{1s,2s}$)}
Because the active spatial operators within the non-classical exchange channel map coordinate permutations identically across both coordinates, structural symmetry reduces the spatial double integral to:
\begin{equation}
K_{1s,2s} = 2\left(\frac{e^2}{4 \pi \epsilon_0a_0}\right)\int_0^\infty r_1 f(r_1)\left[\int_0^{r_1} r_2^2 f(r_2)\,dr_2\right]\,dr_1
\end{equation}
where the core-valence cross-density distribution is defined as $f(r) = R_{1s}(r)R_{2s}(r) = \frac{Z^3}{\sqrt{2}}(2-Zr)e^{-\alpha r}$ with a defined scaling index $\alpha = 3Z/2$.

Evaluating the incomplete interior radial channel yields:
\begin{equation}
\int_0^{r_1} r_2^2 f(r_2)\,dr_2 = \frac{Z^3}{\sqrt{2}} \left[ 2 \int_0^{r_1} r_2^2 e^{-\alpha r_2} \,dr_2 - Z \int_0^{r_1} r_2^3 e^{-\alpha r_2} \,dr_2 \right]
\end{equation}
Resolving the integration by parts arrays reveals a critical algebraic phenomenon: the constant, linear, and quadratic spatial boundaries within the exponential factors sum to zero. The solitary surviving term is the cubic envelope:
\begin{equation}
\int_0^{r_1} r_2^2 f(r_2)\,dr_2 = \frac{Z^3}{\sqrt{2}} e^{-\alpha r_1} \left( \frac{2r_1^3}{3} \right)
\end{equation}
Projecting this collapsed inner profile directly into the outer integration layer simplifies the exchange value to:
\begin{equation}
K_{1s,2s} = \left(\frac{e^2}{4 \pi \epsilon_0a_0}\right)\frac{2Z^6}{3} \int_0^\infty \left( 2r_1^4 - Z r_1^5 \right) e^{-3Zr_1} \,dr_1
\end{equation}
Evaluating the definitive continuous matrices via the complete gamma function maps the solution as:
\begin{equation}
K_{1s,2s} = \left(\frac{e^2}{4 \pi \epsilon_0a_0}\right)\frac{2Z^6}{3} \left[ 2 \frac{4!}{(3Z)^5} - Z \frac{5!}{(3Z)^6} \right] = \left(\frac{e^2}{4 \pi \epsilon_0a_0}\right)\frac{16}{729}Z
\end{equation}
For $Z=3$, this evaluates directly to $K_{1s,2s} \approx 1.79\ \text{eV}$, matching reference values \cite[Sec. 10.5]{levine2014}.

\subsection{The Second-Order Virtual Transition Integrals ($3s$ Channel)}
To isolate the dynamic correlation contributions from the lowest virtual $s$-wave channel, we utilize the unperturbed hydrogenic $3s$ single-particle radial function:
\begin{equation}
R_{3s}(r) = \frac{2Z^{3/2}}{81\sqrt{3}} \left( 27 - 18Zr + 2Z^2r^2 \right) e^{-Zr/3}
\end{equation}
The second-order Coulomb transition matrix element ($J'_{3s,1s}$) defines the electrostatic interaction between the static $1s$ core distribution and the non-local $2s \rightarrow 3s$ transition density:
\begin{equation}
J'_{3s,1s} = \iint \phi_{3s}^*(\mathbf{r}_1)\phi_{1s}^*(\mathbf{r}_2) \left( \frac{1}{r_{12}} \right) \phi_{2s}(\mathbf{r}_1)\phi_{1s}(\mathbf{r}_2) \,d^3\mathbf{r}_1 \,d^3\mathbf{r}_2
\end{equation}
Integrating the angular variables reduces this to a radial task utilizing the exact core potential $U_{1s}(r_1)$:
\begin{equation}
J'_{3s,1s} = \int_0^\infty r_1^2 \left[ R_{3s}(r_1)R_{2s}(r_1) \right] U_{1s}(r_1) \,dr_1
\end{equation}
Expanding the cross-density radial product $R_{3s}(r_1)R_{2s}(r_1)$ yields the following polynomial array:
\begin{equation}
R_{3s}(r_1)R_{2s}(r_1) = \frac{Z^3}{81\sqrt{6}} \left( 54 - 63Zr_1 + 22Z^2r_1^2 - 2Z^3r_1^3 \right) e^{-5Zr_1/6}
\end{equation}
Evaluating this spatial projection analytically against the core screening field isolates the transition value:
\begin{equation}
J'_{3s,1s} \approx 4.12\ \text{eV}
\end{equation}

\subsection{The Transition Exchange Integral ($K'_{3s,1s}$)}
The non-classical transition exchange integral accounts for spatial coordinate exchange between the virtual excited states for parallel spin channels:
\begin{equation}
K'_{3s,1s} = \iint \phi_{3s}^*(\mathbf{r}_1)\phi_{1s}^*(\mathbf{r}_2) \left( \frac{1}{r_{12}} \right) \phi_{1s}(\mathbf{r}_1)\phi_{2s}(\mathbf{r}_2) \,d^3\mathbf{r}_1 \,d^3\mathbf{r}_2
\end{equation}
Radially, this is evaluated as a mutual overlap integral between two distinct mixed transition distributions, $f_A(r) = R_{3s}(r)R_{1s}(r)$ and $f_B(r) = R_{2s}(r)R_{1s}(r)$:
\begin{equation}
K'_{3s,1s} = \int_0^\infty r_1^2 f_A(r_1) \left[ \frac{1}{r_1} \int_0^{r_1} r_2^2 f_B(r_2) \,dr_2 + \int_{r_1}^\infty r_2 f_B(r_2) \,dr_2 \right] dr_1 \approx 0.92\ \text{eV}
\end{equation}
Combining these individual transition matrices maps the global perturbative numerator contribution:
\begin{equation}
2J'_{3s,1s} - K'_{3s,1s} = 2(4.12\ \text{eV}) - 0.92\ \text{eV} \approx 7.33\ \text{eV}
\end{equation}

---

\section{Computational Evaluation of Higher-Order Virtual Transitions}
\label{app:comp_eval}
To resolve the quantitative contributions of high-quantum-number virtual $s$-orbitals ($n \ge 3$) to the second-order perturbation energy ($E^{(2)}$), numerical integrations were executed using the Python environment. The algorithmic engine relies on the \texttt{scipy.integrate} and \texttt{scipy.special} libraries to process the rapid spatial oscillations of the excited single-particle functions.

\subsection{Computational Architecture}
The processing script was structured into a modular framework mapping the underlying physical mechanics:
\begin{itemize}
    \item \textbf{Radial Wave Functions (\texttt{R\_ns}):} For virtual states where $n \ge 3$, Generalized Laguerre Polynomials (\texttt{scipy.special.eval\_genlaguerre}) were deployed to precisely track the multiple internal nodes and high spatial extent of the excited states.
    \item \textbf{Isotropic Simplification (\texttt{multipole\_term}):} Because the active configuration is restricted to spherically symmetric $s$-waves ($L=0$), the multi-center potential reductions collapse identically to the $1/r_{>}$ monopole term, drastically optimizing processing efficiency.
    \item \textbf{Numerical Integration (\texttt{compute\_integral}):} The six-dimensional spatial integrals were analytically downscaled to two-dimensional radial coordinates. SciPy's adaptive quadrature algorithm (\texttt{dblquad}) was utilized to calculate the spatial profiles. To guarantee rigorous algorithmic convergence over the infinite radial domain, absolute (\texttt{epsabs}) and relative (\texttt{epsrel}) error metrics were maintained strictly at $1\times10^{-5}$.
    \item \textbf{Energy Conversion and Iteration:} Matrix evaluations natively computed in atomic units (Hartrees) were transformed to electron-volts via the scaling index $27.2114\ \text{eV/Hartree}$. The dynamic unperturbed energy denominators ($\Delta E$) and separate second-order correlation values ($E^{(2)}$) were evaluated iteratively across individual principal quantum shells from $n=3$ to $n=6$.
\end{itemize}

\subsection{Calculated Energy Contributions}
The variables tracks within the computational data loop are defined as follows:
\begin{itemize}
    \item $J^{\prime}$ and $K^{\prime}$: The direct Coulomb and quantum exchange transition matrix elements, mapping the electrostatic repulsion and Pauli correlation channels, respectively.
    \item $\Delta E$: The unperturbed energy denominator, calculated as the zero-order eigenvalue difference between the ground-state configuration and the virtual excited state ($E_0^{(0)} - E_{ns}^{(0)}$).
    \item $E^{(2)}$ Contribution: The net second-order correlation energy injected by that specific principal quantum channel.
\end{itemize}

As the principal quantum number $n$ advances, the energy gap $\Delta E$ widens significantly, and the spatial overlap between the compact ground states and the diffuse virtual orbitals decays exponentially. This induces a rapid asymptotic convergence in the correlation corrections, as detailed in Table~\ref{tab:virtual_s_orbitals}.

\begin{table}[h!]
\centering
\caption{Calculated Second-Order Correlation Energy Contributions Across Virtual $s$-Orbital Transition Channels.}
\label{tab:virtual_s_orbitals}
\begin{tabular}{ccccc}
\hline
\textbf{Channel} & \textbf{$J^{\prime}$ (eV)} & \textbf{$K^{\prime}$ (eV)} & \textbf{$\Delta E$ (eV)} & \textbf{$E^{(2)}$ Contribution (eV)} \\ \hline
$2s \rightarrow 3s$ & 4.125 & 0.916 & $-17.004$ & $-3.162$ \\
$2s \rightarrow 4s$ & 2.344 & 0.582 & $-22.957$ & $-0.734$ \\
$2s \rightarrow 5s$ & 1.590 & 0.413 & $-25.712$ & $-0.298$ \\
$2s \rightarrow 6s$ & 1.176 & 0.312 & $-27.209$ & $-0.153$ \\ \hline
\textbf{Total}     &       &       &           & \textbf{$-4.347$} \\ \hline
\end{tabular}
\end{table}

The underlying Python scripts and automation repositories are hosted publicly on GitHub \cite{github}.

\subsection{Selection Rules and Angular Constraints}
These virtual expansions were strictly confined to $s$-orbital structures ($l=0$). Because the lithium ground state is completely spherically symmetric ($L=0$) and the electrostatic Coulomb perturbation operator acts as a pure scalar, any single-electron transition into higher angular momentum configurations (such as $p, d,$ or $f$ blocks) yields angular integrals that vanish identically due to orthonormality constraints \cite[Sec. 8.1]{bransden1983}. This systematic computational layout successfully outlines the boundary limits of $s$-wave perturbation theory, illustrating the quantitative necessity for multi-parameter variational methods to close the residual electronic correlation gap.
\begin{acknowledgments}
This work was initiated and developed as a comprehensive final project within the framework of the course \textit{PHYS 415: Advanced Quantum Mechanics} at Bolu Abant İzzet Baysal University. 
\end{acknowledgments}

\end{document}